\documentclass[floatfix, aps, prd, a4paper, 10pt, showpacs, superscriptaddress, nofootinbib, reprint]{revtex4-1}
\tighten
\usepackage{color}
\usepackage{xcolor}
\usepackage{amsmath,amssymb,graphicx}
\usepackage{wasysym}
\usepackage{bm,lipsum}
\usepackage{times}
\usepackage{multirow}
\usepackage{microtype}
\usepackage{comment}
\usepackage[utf8]{inputenc}
\usepackage{booktabs}
\usepackage{subfigure}
\usepackage[normalem]{ulem}
\usepackage[varg]{txfonts}
\usepackage{bm,graphics,graphicx,epsfig,color,ulem}
\usepackage{array}
\usepackage{tabularx}
\usepackage[colorlinks, pdfborder={0 0 0}]{hyperref}
\usepackage{bm}

\definecolor{LinkColor}{rgb}{0.75, 0, 0}
\definecolor{CiteColor}{rgb}{0, 0.5, 0.5}
\definecolor{UrlColor}{rgb}{0, 0, 0.75}
\definecolor{maroon}{rgb}{0.5, 0.0, 0.0}
\definecolor{mint}{rgb}{0.24, 0.71, 0.54}
\definecolor{violet}{rgb}{0.62, 0.0, 1.0}
\hypersetup{linkcolor=LinkColor}
\hypersetup{citecolor=CiteColor}
\hypersetup{urlcolor=UrlColor}
\allowdisplaybreaks
\newcolumntype{C}{>{\arraybackslash}m{6cm}}

\newcommand{\aplus} {\texttt{HLVKI+}}
\newcommand{\voy} {\texttt{VK+HLIv}}
\newcommand{\et} {\texttt{HLKI+E}}
\newcommand{\ce} {\texttt{VKI+C}}
\newcommand{\EC} {\texttt{KI+EC}}
\newcommand{\ECS} {\texttt{ECS}}

\newcommand{\lamt}{$\tilde{\Lambda}$~}

\newcolumntype{?}{!{\vrule width 1pt}}

\begin{document}

\title{The Accuracy of Neutron Star Radius Measurement with the Next Generation of Terrestrial Gravitational-Wave Observatories}
\author{Rachael Huxford}
\affiliation{Institute for Gravitation and the Cosmos, Department of Physics, Pennsylvania State University, University Park, PA, 16802, USA}

\author{Rahul Kashyap}
\affiliation{Institute for Gravitation and the Cosmos, Department of Physics, Pennsylvania State University, University Park, PA, 16802, USA}

\author{Ssohrab Borhanian}
\affiliation{Theoretisch-Physikalisches Institut, Friedrich-Schiller-Universit\"at Jena, 07743, Jena, Germany}
\affiliation{Institute for Gravitation and the Cosmos, Department of Physics, Pennsylvania State University, University Park, PA, 16802, USA}

\author{Arnab Dhani} 
\affiliation{Max Planck Institute for Gravitational Physics (Albert Einstein Institute), Am Mühlenberg 1, Potsdam 14476, Germany}
\affiliation{Institute for Gravitation and the Cosmos, Department of Physics, Pennsylvania State University, University Park, PA, 16802, USA}

\author{Ish Gupta}
\affiliation{Institute for Gravitation and the Cosmos, Department of Physics, Pennsylvania State University, University Park, PA, 16802, USA}

\author{B. S. Sathyaprakash}
\affiliation{Institute for Gravitation and the Cosmos, Department of Physics, Pennsylvania State University, University Park, PA, 16802, USA}
\affiliation{Department of Astronomy \& Astrophysics, Pennsylvania State University, University Park, PA, 16802, USA}
\affiliation{School of Physics and Astronomy, Cardiff University, Cardiff, UK, CF24 3AA}

\begin{abstract}
    In this paper, we explore the prospect for improving the measurement accuracy of masses and radii of neutron stars. We consider imminent and long-term upgrades of the Laser Interferometer Gravitational-Wave Observatory (LIGO) and Virgo, as well as next-generation observatories---the Cosmic Explorer and Einstein Telescope. We find that neutron star radius with single events will be constrained to within roughly 500 m with the current generation of detectors and their upgrades. This will improve to 200 m, 100 m and 50 m with a network of observatories that contain one, two or three next-generation observatories, respectively. Combining events in bins of 0.05 $M_\odot$ we find that for stiffer (softer) equations-of-state like ALF2 (APR4), a network of three XG observatories will determine the radius to within 30 m (100 m) over the entire mass range of neutron stars from  $1\,M_\odot$ to $2.0\,M_\odot$ ($2.2\,M_\odot$), allowed by the respective equations-of-state. Neutron star masses will be measured to within 0.5\% with three XG observatories irrespective of the actual equation-of-state. Measurement accuracies will be a factor of 4 or 2 worse if the network contains only one or two XG observatories, respectively, and a factor of 10 worse in the case of networks consisting of Advanced LIGO, Virgo KAGRA and their upgrades. Tens to hundreds of high-fidelity events detected by future observatories will allow us to accurately measure the mass-radius curve and hence determine the dense matter equation-of-state to exquisite precision.
\end{abstract}

\maketitle

\section{Introduction and Background}
An outstanding problem in nuclear astrophysics is the  equation-of-state of neutron star (NS) cores, believed to contain matter at several times the nuclear saturation density \cite{Lattimer:2015nhk, Ozel:2016oaf, Baym:2017whm}: near the core the density reaches 4 to 6 times the nuclear saturation density and in the outer core it would be twice the nuclear saturation density.
This makes them the densest objects anywhere in the Universe. Decades after their discovery, the radii of neutron stars are still uncertain{\footnote{Note that some authors, who claim a 5\% uncertainty in the radius, are quoting one-sided, one-$\sigma$ credible intervals. The 10\% to which we refer corresponds to a two-sided, 90\% credible interval, which is the standard in LIGO-Virgo Collaboration publications.}} by about $\sim 10\%$ \cite{Raaijmakers:2019dks, Raaijmakers:2021uju,Ayriyan:2021prr, Essick:2020flb, Hu:2020ujf, Raithel:2019uzi,Miller:2021qha}, and the composition of their dense cores likely depends on the neutron star mass and could be composed of hadrons or deconfined quarks \cite{Baym:2017whm,Han:2019bub, Li:2021crp, Christian:2018jyd, Montana:2018bkb}. Indeed, it is not clear whether the matter at such densities undergoes a phase transition from a hadronic phase to quark-gluon plasma and the critical neutron star mass and temperature at which the transition might occur \cite{Baym:2017whm, Han:2019bub,Li:2021crp, Christian:2018jyd, Montana:2018bkb, Li:2021crp, Li:2018ayl, Blaschke:2020vuy}.

Neutron stars in binaries are studied either as radio pulsars or X-ray sources and both have helped in our understanding of the structure of neutron stars \cite{Postnov:2014tza,Steiner:2011ft, Catuneanu:2013pz, Guver:2010td, Guver:2008gc, Lattimer:2013hma, Poutanen:2014xqa}. The Neutron Star Interior Composition Explorer (NICER) space observatory is providing precision X-ray data on neutron stars \cite{Gendreau2016-oe}. Precise general relativistic modeling of the X-ray pulsation of neutron stars has been used to constrain their masses and radii as well as the equation-of-state (EOS) of their dense cores \cite{Miller:2019cac, Raaijmakers:2019qny, Bogdanov:2019ixe, Bogdanov:2019qjb, Watts:2019lbs, Raaijmakers:2019dks, Miller:2021qha}. The best-measured NICER radius errors are about 1 km.

At the same time, advances in gravitational-wave observations from merging neutron stars are allowing new approaches to resolve this puzzle. Indeed, the detection of binary neutron stars (BNSs) \cite{ LIGOScientific:2014pky, VIRGO:2014yos, LIGOScientific:2018hze, LIGOScientific:2017vwq, LIGOScientific:2018mvr} and neutron star-black hole binaries (NSBHs) \cite{LIGOScientific:2020aai} by the Laser Interferometer Gravitational-Wave Observatory (LIGO) and Virgo has opened up a new and independent window for exploring neutron stars. Gravitational waves emitted in the final tens of milliseconds of the inspiral and coalescence of BNSs can be used to explore the composition and EOS of dense matter in neutron star cores \cite{Lai:1993pa, Cutler:1994ys, Kokkotas:1999bd,  Flanagan:2007ix, Hinderer:2009ca}. Encoded in the phase evolution of the waves is the (dimensionless) \emph{tidal deformability} $\Lambda_{1,2}$ of the two stars, which is a measure of the quadrupole deformation imparted on the stars due to the tidal field of their companions. The leading order finite size effect in the post-Newtonian (PN) approximation of the waves' phase evolution is a highly sub-dominant effect. In terms of the post-Newtonian expansion parameter $(v/c) < 1,$ it is, in fact, an order ${\cal O}(v/c)^{10}$ effect beyond the dominant quadrupole term \cite{Flanagan:2007ix, Hinderer:2007mb, Hinderer:2009ca, Vines:2011ud}, yet it is significant when the instantaneous gravitational-wave frequencies are $\sim 100$ Hz or greater ($v/c \sim 0.16$ or larger) for a typical BNS system comprising a pair of 1.4 $M_\odot$ companions \cite{Harry:2021hls}. 

The tidal deformability goes as the inverse fifth power of the star's compactness, i.e. $\Lambda_k \propto [Gm_k/(c^2R_k)]^{-5},$ $k=1,2,$ where $m_k$ and $R_k$ are the masses and radii of the companion stars in a binary system \cite{Hinderer:2007mb, Favata:2013rwa}. Matched filtering the data with gravitational-wave templates calibrated to numerical relativity simulations \cite{Hinderer:2016eia, Dietrich:2017feu, Dietrich:2017aum, Dietrich:2018uni, Dietrich:2019kaq, Nagar:2018zoe,Henry:2020pzq,Henry:2020ski} of BNS mergers can be used, in principle, to measure the tidal deformabilities of the companions, in addition to their masses\footnote{Neutron stars in merging binaries are not expected to have large spins. Consequently, the only intrinsic parameters that we will consider in this paper are the masses and the tidal deformabilities.}. In practice, however, it is not possible to accurately measure the individual tidal deformabilities, but only a certain linear combination of the two called \emph{effective tidal deformability} $\tilde{\Lambda}$, defined by:
    \begin{equation}
    \tilde{\Lambda} 
    = \frac{16}{13(1+q)^5} \left[ \left(1+12 q\right) \Lambda_1 + q^4 \left( 12+q\right) \Lambda_2 \right]
    \label{eq:lambdatilde}
     \end{equation}
where  $q\equiv m_2/m_1 \le 1$ is the mass ratio \cite{Flanagan:2007ix, Vines:2011ud, Favata:2013rwa, Wade:2014vqa}.  Although the dominant tidal effect, which depends only on $\tilde\Lambda,$ is measured accurately, the PN correction, required to measure the individual tidal deformabilities, cannot be inferred with any accuracy. This is because of two reasons: On the one hand, it is a higher order PN correction, an ${\cal O}(v/c)^{12}$ effect, compared to the dominant quadrupole term and, on the other hand, the PN correction vanishes for binaries with comparable masses. In fact, the tidal PN correction depends on $\delta\tilde\Lambda$ defined by:
    \begin{eqnarray}
    \delta \tilde{\Lambda} & = & \sqrt{1-4 \eta} \left( 1-\frac{13272}{1319} \eta +\frac{8944}{1319} \eta^2 \right) \left ( \frac{\Lambda_{2}+\Lambda_1}{2} \right ) \nonumber \\ 
     & + &  \left( 1-\frac{15910}{1319} \eta +\frac{32850}{1319} \eta^2 +\frac{3380}{1319} \eta^{3} \right)  \left ( \frac{\Lambda_{2}-\Lambda_1}{2} \right ),
     \label{eq:deltalambda}
    \end{eqnarray}
where $\eta\equiv m_1m_2/(m_1+m_2)^2 = q/(1+q)^2$ is the symmetric mass ratio. For BNS systems in general, companion masses are similar, and hence $q\simeq 1$ and $\Lambda_1 \simeq \Lambda_2,$ giving $\tilde\Lambda \simeq \Lambda_{1,2}$ and hence $\delta\tilde\Lambda \simeq 0.$ Additionally, the tidal deformability of a neutron star depends not only on its mass, but also the (unknown) EOS. For neutron stars of $1.4 M_\odot$ and over a wide range of equations-of-state (EOSs), typical values are $\Lambda_{k} \sim 200$--2000 \cite{Favata:2013rwa}. While the first post-Newtonian correction is already sub-dominant as a sixth post-Newtonian order effect compared to the leading order quadrupole \cite{Favata:2013rwa}, this range of $\Lambda_{k}$ also results in the term being
at least two orders of magnitude smaller compared to the leading order tidal term. These effects combined make the term difficult to measure. Consequently, only the leading order tidal term, is readily available, making it necessary to supplement gravitational-wave observations with other input in order to infer the individual tidal deformabilities and the radii of neutron stars. Several such approaches have been proposed in the literature and applied to GW170817 \cite{LIGOScientific:2018hze, Riley:2018ekf}. 

The BNS coalescence event \textsc{GW170817}, at $\sim 40$~Mpc and a signal-to-noise ratio (SNR) of 33, provided the first opportunity to constrain the tidal deformabilities from gravitational-wave observations, and hence the radii, of neutron stars \cite{LIGOScientific:2017vwq, LIGOScientific:2018hze, LIGOScientific:2018mvr}. Theoretical models of the EOS of neutron stars are plenty and varied and they allow tidal deformabilities in the range of $10 \lesssim \Lambda_{1,2} \lesssim 10000$ \cite{Hinderer:2007mb, Hinderer:2009ca}, depending on the mass, being larger for lighter neutron stars and stiffer EOSs. Analysis of the event \textsc{GW170817} found that the 90\% credible range of the companion masses were $1.36\,M_\odot \le m_1 \le 1.89\,M_\odot$ for the primary and $1.00\,M_\odot \le m_2 \le 1.36\,M_\odot$ for the secondary \cite{LIGOScientific:2017vwq}, the effective tidal deformability had a 90\% credible upper bound of  $\tilde\Lambda\lesssim 600$ and the radius was constrained to be 
 $R_{1} = 11.9 ^{+1.4}_{-1.4}$ km 
\cite{LIGOScientific:2018hze,De:2018uhw}. Unfortunately, the second BNS event \textsc{GW190425} \cite{GW190425} was farther and had a significantly lower SNR than \textsc{GW170817} and did not yield tighter constraints on the tidal deformability on its own.
 
 However, constraints have also been derived by combining LIGO-Virgo results of \textsc{GW170817} and \textsc{GW190425} with additional observations. Including NICER observations \cite{Raaijmakers:2019dks, Raaijmakers:2021uju, Miller:2019cac, Raaijmakers:2019qny, Bogdanov:2019ixe, Bogdanov:2019qjb, Watts:2019lbs, Miller:2021qha} bound the radius of a $1.4\,M_\odot$ neutron star to the range $R_{1.4} = 12.33 ^{+0.76}_{-0.81}$ km. Likewise, combining nuclear physics experiments and gravitational-wave data has found $R_{1.4} = 11.0 ^{+0.9}_{-0.6}$ km \cite{Capano:2019eae}, and  $R_{1.4} = 12.75 ^{+0.42}_{-0.54}$ km \cite{Biswas:2021yge} while combining data from GW170817, its companion gamma-ray burst GRB170817A, and subsequent kilonova AT2017gfo, the same radius 
 was determined to an accuracy of less than about a km at 90\% credible interval \cite{Breschi:2021tbm}. However, see \citealt{Vinciguerra:2023qxq} for sensitivity of NICER results on model hypotheses.

The planned upgrades of LIGO and Virgo, the addition of observatories currently under construction,  KAGRA~\cite{KAGRA:2018plz} in Japan and LIGO-Aundha in India \cite{Saleem:2021iwi}, and new, longer-arm facilities that are currently being conceived, have the potential to make new discoveries of both sources and science. In this study, we explore the accuracy with which future observatories are able to measure the radii of neutron stars, an important step in constraining their equation of state.  The networks considered in this work include the imminent upgrade of LIGO and Virgo over the next five years called A+~\cite{LIGOScientific:2014pky,VIRGO:2014yos}, the \emph{Voyager} upgrade to LIGO detectors that would be possible within the next ten years \citep{Adhikari2023-ar}, and the next-generation (XG) observatories such as the Einstein Telescope (ET) \cite{Punturo:2010zza, Sathyaprakash:2012jk, Maggiore:2019uih} or Cosmic Explorer (CE) \cite{Reitze:2020} that are expected to operate in the mid-2030s in tandem with the fully upgraded versions of current observatories. Given the rate of BNS mergers as determined by the events \textsc{GW170817} and \textsc{GW190425}, we expect the future observations to constrain the neutron star radius to within 600 m (A+ generation), 400 m (Voyager generation), 200 m (one XG observatory) and $<100$ m (two or more XG observatories), with the high-fidelity events observed by the respective networks of observatories. At the same time, neutron star masses will be measured to better than 10\%, 5\%, 3\% and 0.5\% \cite{Puecher:2023twf}. The mass-radius relation is a proxy to the EOS of ultra-dense matter in neutron-star cores that will be tightly constrained with high-precision measurements of the masses and radii with future networks of gravitational-wave observatories (see, e.g., \cite{Pacilio:2021jmq}).

When combining information from a multiple set of events it is necessary to employ a population model for the observed sources in addition to the unknown equation of state. For binary neutron stars, the population model will involve the astrophysical distribution of neutron star masses (or, equivalently, the neutron star central densities), the pairing probability as a function of the total mass and mass ratio and the distribution of neutron star spins. Moreover, gravitational-wave detectors and the analysis pipeline used to detect binary neutron stars have selection effects. For example, it is easier to detect equal-mass systems compared to mass-asymmetric systems of the same total mass. Likewise, binaries with a larger total mass produces a larger signal-to-noise ratio compared to a binary of smaller total mass but the same mass ratio. Bayesian inference of the source parameters for a single event will also be affected by the unknown hyper parameters of the population model since the posterior distribution depends on the assumed prior model. Thus, one has to simultaneously determine the population model and the EOS. For the EOS, this means one has to marginalize over the population model. Additionally one must also account for the selection effects to assure that the model selection of  EOS is unbiased.\\
We are ignoring these effects in this work since our Fisher matrix approach currently does not allow for the inclusion of systematic biases. We also envisage that in the XG era the selection effects would have been better understood. Our goal, instead, is to provide the statistical uncertainty that we expect in the determination of the EOS. We are currently in the process of preparing a mock data challenge for XG observatories. The mock data challenge will allow us to address the aforementioned issues.\\
We also note that the estimation of intrinsic source masses requires the use of a cosmological model. Since we detect BNS events to a significant cosmological distance, cosmological parameters must be inferred together with the parameters of a BNS event \citep{Ghosh:2022muc, Chatterjee:2021xrm}. As explained in Sec.\,\ref{sec:cosmology}, we find that the bias introduced due to an unknown cosmological model is negligible.

The rest of the paper is organized as follows.  In Sec.\,\ref{sec:injections} we describe the cosmic BNS population used in this study together with the distribution of companion masses, the merger rate and its variation with redshift and the waveform model used.  This is followed by a brief summary of detector networks considered in Sec.\,\ref{sec:future}, focusing on the efficiency of the networks in detecting BNS systems.    In Sec.\,\ref{sec:measurement} we present the capabilities of the different observatories in characterizing the source properties.  We describe in  Sec.\,\ref{sec:tidal deformability and radius} the method to infer the  radii of neutron stars from the measurement of effective tidal deformability using a set of EOS independent universal relations with corrections and how we combine the results from a population to obtain joint bounds. In Sec.\,\ref{sec:results} we present the application of the methods to events expected to be observed in detector networks considered in this study. The results are obtained by combining radius  measurements of a small sub-population of observed events: either the loudest 100 events or the 100 events for which tidal deformability is best measured, to infer the radii of neutron stars. A summary of the results and conclusions is presented in  Sec.\,\ref{sec:conclusions}.

\section{Neutron Star Population and Waveform Model}
\label{sec:injections}
    In this section, we describe the neutron star population and the waveform approximations used in the study. We begin by recalling how the redshift dependence of the merger rate is computed using the observed star formation rate as a function of redshift as a proxy for the redshift evolution of the rate. The redshift dependence is not exactly the same as the star formation rate since binaries that form from stars only merge after a certain time delay, which is essentially the gravitational radiation back reaction timescale.  This is followed by a summary of the distribution of neutron star masses used in the study. We conclude the section with a description of the waveform model used, which is built upon the point-particle approximation but includes finite-size tidal effects with the waveform model parameters calibrated to hydrodynamical numerical relativity simulations of BNS mergers.

\subsection{BNS Merger Rate}
The merger rate density $r_0$ in the local Universe (i.e., at zero redshift) inferred from LIGO-Virgo observations of BNS coalescences during the second and third observing runs is $r_0=10-1700\,\mbox{ yr$^{-1}$ Gpc$^{-3}$}$ \cite{KAGRA:2021duu}. The two BNS events observed during this period, \textsc{GW170817} and \textsc{GW190425}, were localized to luminosity distances of $40\,\rm Mpc$ and $159\,\rm Mpc,$ respectively, and corresponding redshifts of $z\simeq 0.01$ and $z\simeq 0.036.$ Thus, the LIGO-Virgo rate is essentially the \emph{local} merger rate density, i.e. at redshift $z=0.$ In this work, we will consider mergers up to a redshift of $z=1.$ The merger rate density over this redshift range is expected to increase since the rate of star formation $\psi(z)$, from which compact binaries form, increases with redshift up to about $z=2$ \cite{Madau:2014bja}.

To model the variation of the merger rate with redshift, we assume that it follows the star formation rate except that a binary that forms at redshift $z_1$ merges at redshift $z<z_1.$ This is because there could be a significant time-delay $t_d$ between the binary's initial formation and eventual merger as driven by gravitational radiation back reaction. The time delay $t_d$ for a specific binary depends on a number of astrophysical processes that take place between the formation of the companion stars, their common evolution, and survival following supernova kicks they receive. Therefore, $t_d$ will not be the same for every binary and the time delay distribution is not well known either due to the complexity of how the progenitors of compact binaries evolve. However, making reasonable assumptions about the intervening processes, i.e. neutron stars form with no delay after the formation of their progenitor stars, their orbit decays due to the emission of gravitational waves only, and the semi-major axis of their orbit follows a uniform in log-space  distribution, $t_d$ follows the distribution $P(t_d)\propto 1/t_d$ \cite{Beniamini:2019iop, Greggio:2020vyk}. Thus, the merger rate density in the source's frame\footnote{In what follows lower case letters are used to denote the merger rate densities while capital letters are used to denote the merger rates.}  $r_z(z)$ is given by:
\begin{equation}
    r_z(z) = A \int_{t_d^{\rm min}(z)}^{t_d^{\rm max}(z)} \psi(z-t_d(z))\,P(t_d(z))\,\frac{dt_d}{dz}\, dz,
\end{equation}
where a subscript $z$ is included to clarify that $r_z(z)$ is the rate density with respect to an observer at $z,$ 
$t_d^{\rm min}$ and $t_d^{\rm max}$ are the minimum and maximum time delays, $A$ is a normalization constant (see below), and  $\psi(z)$ denotes the star formation rate (whose dimensions are not important to us but only its dependence on redshift). For $\psi(z)$ we use the fit proposed in Ref.~\cite{Vangioni:2014axa}: 

\begin{equation}
\psi(z) \propto \frac{a\exp{(b(z-z_m))}}{a-b+b \exp{(a(z-z_m))}}
\end{equation}
where $a=2.8$, $b=2.46$, and $z_m=1.72$.
For the minimum time-delay we use $t_d^{\rm min}=0.2$ Gyr and for the maximum we use $t_d^{\rm max} = 10$ Gyr. The normalization constant $A$ is determined so that this expression is consistent with the local rate density, i.e. $r_0(z=0)=r_0$. The merger rate, $r_z(z)$, peaks at a slightly lower redshift than $\psi(z)$ because of the time-delay. The dependence of the cosmic time $t$ on redshift is determined by the Planck 2015 Cold Dark Matter cosmology:
\begin{equation}
\frac{dt}{dz} = \frac{1}{H_0 (1+z)\,\sqrt{\Omega_\Lambda + \Omega_M(1+z)^3}},
\end{equation}
with the Hubble constant $H_0=69.6\,\rm km\, s^{-1}\,Mpc^{-1},$ $\Omega_\Lambda=0.714$ and $\Omega_M=0.286.$


Next, the merger rate (as opposed to rate density) in a redshift interval $dz$ is given by:
\begin{equation}
    dR_z(z) = r_z(z)\,\frac{dV}{dz}\,dz
    \label{eq:dRz}
\end{equation}
where $dV=(dV/dz)\,dz$ is the comoving volume element corresponding to redshift range $dz.$ To convert this to the rate as measured by an observer at $z=0$ we must divide by $(1+z)$ to take into account the redshift of the rate due to cosmological expansion: $dR_0(z)=dR_z(z)/(1+z).$ The cumulative merger rate is given by:
\begin{equation}
    R(z) = \int_0^z dR(z') = \int_0^z \frac{r_z(z')}{(1+z')}\,\frac{dV}{dz'}\,{\rm d}z'.
    \label{eq:merger rate}
\end{equation}

Within $z=1$ the merger rate is about $\sim 10^5$ per year. Not all of these mergers would be detectable by a gravitational-wave detector (or a network) but only a certain fraction depending on its sensitivity which we will discuss in Sec.\,\ref{sec:future}.

\subsection{Waveform Models and Mass Distribution}
\label{sec:waveforms}
In order to characterize the capability of various detector networks to measure the tidal deformability and the companion masses, it is important to choose an appropriate waveform model that includes the relevant physical effects.
As in the case of binary black holes, BNS waveforms are based on approximate solutions to Einstein equations. They include the dominant tidal effects and incorporate additional parameters in the phase evolution which are calibrated by matching the analytical solution against numerical relativity simulations.
We chose the frequency-domain phenomenological waveform model \textsc{IMRPhenomPv2NRTidalv2} \cite{Dietrich:2017aum,Dietrich:2018uni,Dietrich:2019kaq} for the generation of simulated signals as well as templates for Fisher-matrix based inference. This model is based on the \textsc{IMRPhenomPv2} BBH waveform \cite{Hannam:2013oca,Khan:2015jqa} with tidal effects up to 7.5 post-Newtonian order (or to ${\cal O} (v/c)^{13}$ beyond the leading quadrupole term), making it appropriate for use in BNS analysis. An earlier version of this waveform was used for the analysis of \textsc{GW170817} \cite{LIGOScientific:2018cki}.

The waveform model takes as input the intrinsic masses of the companions and their tidal deformabilities. In this paper, the companion masses are drawn from a uniform distribution over a range of masses whose lower limit is $1\,M_\odot$ and the upper limit is the maximum allowed by the EOS used in the simulation (see below): $m_1, m_2 \sim U(1\,M_\odot,\,M^{\rm EOS}_{\rm max}).$ 
Although the masses of neutron stars in the Milky Way seem to be concentrated around $1.4\,M_\odot,$ there is \emph{a priori} no physical reason to assume that this is the preferred value in other galaxies. Theoretically, neutron star masses are allowed to be as large as $2.9\,M_\odot$ \citep{Godzieba:2020bbz}, although the largest measured masses tend to be significantly lower. The heaviest neutron stars among astronomical observations are in range $2.01$--$2.35\, M_\odot$ \citep{Antoniadis:2013pzd, Fonseca:2021wxt, Romani:2022jhd}, while from gravitational-wave observations the companion masses in BNS systems are as large as $1.6\,M_\odot,$ and $1.4\,M_\odot$ in the case of \textsc{GW170817}, and $1.9\,M_\odot,$ and $1.7\,M_\odot$ in the case of \textsc{GW190425}. Neutron-star masses in neutron star-black hole systems \textsc{GW200105} and \textsc{GW20015} \cite{LIGOScientific:2021qlt} are both $2.2\,M_\odot.$ In this small population there seems to be no preference for the Galactic value of $\sim 1.4\,M_\odot$ and it would be more prudent to assume a wider range for the mass distribution. We have chosen the widest range allowed by the model EOSs considered in this paper.

We assume, however, that the dimensionless spin magnitudes of neutron stars are negligible.  The fastest-spinning Galactic pulsar has a rotational frequency of just over 700 Hz. Its dimensionless spin angular momentum is still roughly $a=cI2\pi\omega/Gm^2 \simeq 0.4$---far smaller than the maximum spin neutron stars could, in principle, have; here $I$ is the principal moment-of-inertia of the star (roughly equal to $\tfrac{2}{5}mR^2,$ where $m$ and $R$ are the neutron star's mass and radius, respectively), and $\omega$ is its spin angular frequency. Neutron star spins in other galaxies could be far greater than those in the Milky Way but the waveform models that are currently available are calibrated against numerical relativity simulations of BNSs with small spins (dimensionless spin, $\chi < 0.1$) \cite{Dietrich:2017aum,Dietrich:2018uni,Dietrich:2019kaq}. 

In addition to masses and spins, we also have to specify the distance to the source, its orientation relative to the detector frame, and its position in the sky. Sources are assumed to follow the redshift distribution determined by Eq.\,(\ref{eq:dRz}) and uniformly distributed over the angular parameters describing the sky position and orientation of the binary.

Given the mass, the radius of the neutron star is calculated for a given EOS by solving the Tollmann-Oppenheimer-Volkoff (TOV) equations \cite{Oppenheimer:1939ne, Tolman:1939jz}. In practice, this is computationally too expensive since our simulations have to deal with hundreds of thousands of systems. Thus, it is more practical to solve the TOV equations to obtain radii for a set of masses and then use an interpolating function to find the radius for an arbitrary value of mass. We have confirmed that the fractional difference in the radius, for a given mass, obtained from  numerical solution to the TOV equation and the interpolating function are below 0.1\% over the full range of neutron star masses allowed by the EOS.

\begin{figure}[tb!]
    \centering
    \includegraphics[width=1.0\linewidth]{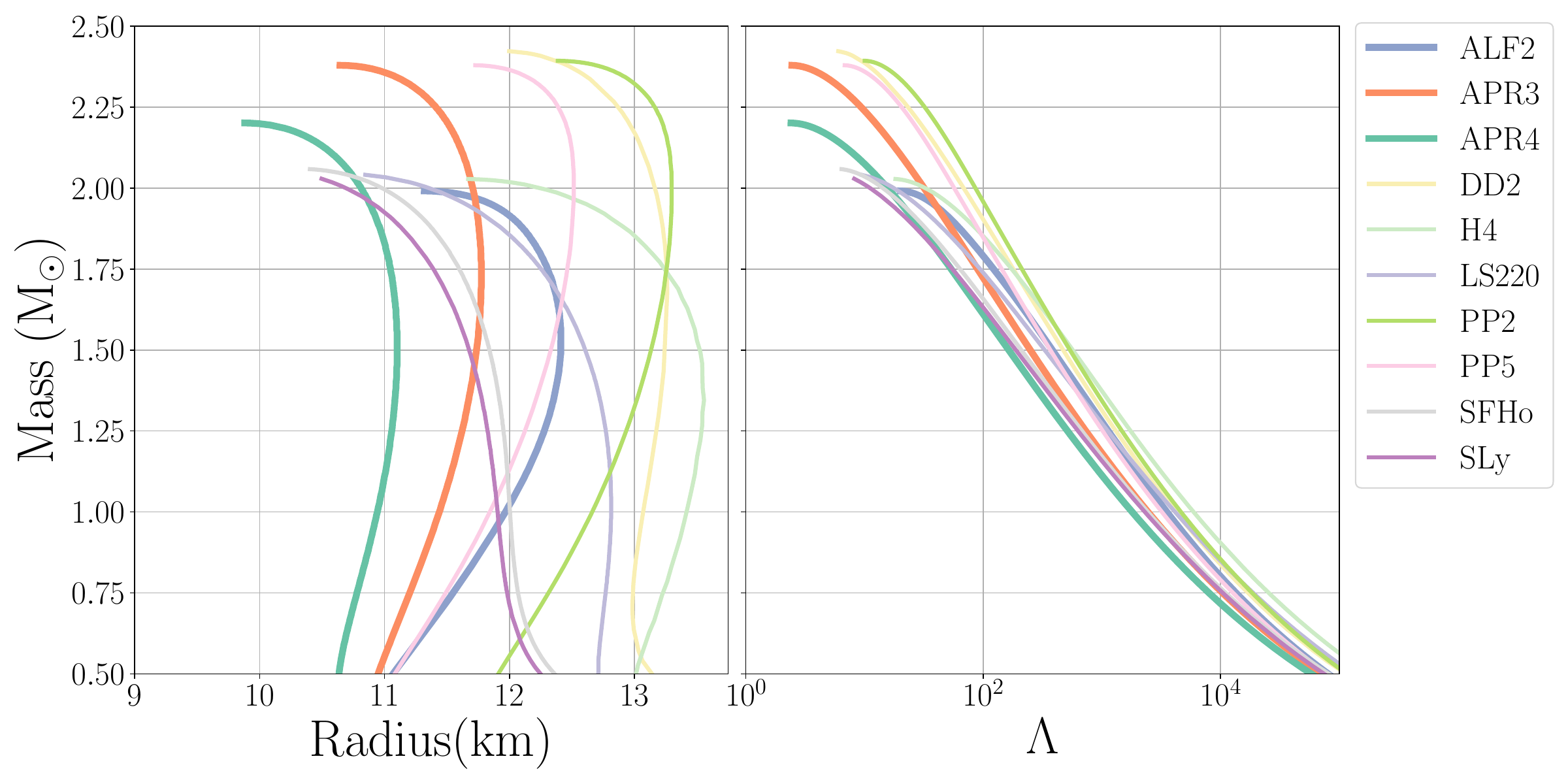}
    \caption{ Mass-Radius curves for EOS used in this paper. Please note that our choice of three injections EOS here (ALF2, APR3, APR4 shown with thicker lines) are motivated by the conservative constraint on $\Lambda_{1.4} <800$ as put forward by \cite{LIGOScientific:2018cki}. We also consider the fact that these three EOS covers a significant range in the maximum masses while the inclusion of addtional seven EOS provide good coverage of the rest of the mass radius space.}
    \label{fig:EOS_mr}
\end{figure}

We consider three EOS used for injection, and an additional seven EOS used for reference that are still allowed by X-ray and gravitational-wave data: the injection set of ALF2, APR3, APR4 and reference set of DD2, H4, S220, PP2, PP5, SFHo, and SLy. We then plot the corresponding mass-radius curves in Fig.\,\ref{fig:EOS_mr}. ALF2 (APR4) represents a stiffer (softer) EOS allowing for larger (smaller) radii, while APR3 allows intermediate radii. The reference EOS then provide good coverage of the mass-radius parameter space between the three, allowing for stronger model discrimination tests with our methods. Given the mass $M_i$ and radius $R_i$, the dimensionless tidal deformability is computed using the expression:
\begin{equation}
    \Lambda_i = \frac{2\, k_2(R_i)}{3} \left ( \frac{c^2R_i}{Gm_i} \right )^5,
\end{equation}
where $k_2(R)$ is the tidal Love number, which also depends on the radius of the neutron star and is fixed for a given mass and EOS \cite{Hinderer:2007mb}.

\section{Future Observatories and Their Reach for the Binary Neutron Star Population}
\label{sec:future}
Advanced LIGO (aLIGO) and Advanced Virgo (AdV) are currently taking data and are expected to reach their design sensitivity goals (see Fig.\,\ref{fig:sensitivities}) in late 2023\footnote{\label{fo:igwn}For up to date schedule of the runs see \url{https://rtd.igwn.org/projects/userguide/en/latest/capabilities.html}.}~\cite{LIGOScientific:2014pky}.  At that sensitivity, the network of LIGO-Hanford, LIGO-Livingston, and Virgo (HLV)~\cite{VIRGO:2014yos} could detect $\sim 40$ BNS mergers per year from within a distance of about 400 Mpc. Both projects have concrete plans to upgrade their sensitivity over a period of two years, which we will refer to as the HLV+ network, enhancing the detection rate by about a factor $\sim 5$ by about 2027\textsuperscript{\ref{fo:igwn}}.

\subsection{Upgrades and New Facilities}
The Japanese KAGRA detector, currently being commissioned, and LIGO-India are  expected to join the HLV+ network over the 2020-2030 decade and the five detectors would be together referred to as the HLVKI+ network. HLV+ and HLVKI+ networks will begin to observe events with SNRs large enough to facilitate accurate measurement of the tidal deformability. 

\begin{figure}[htb]
    \centering
    \includegraphics[width=0.45\textwidth]{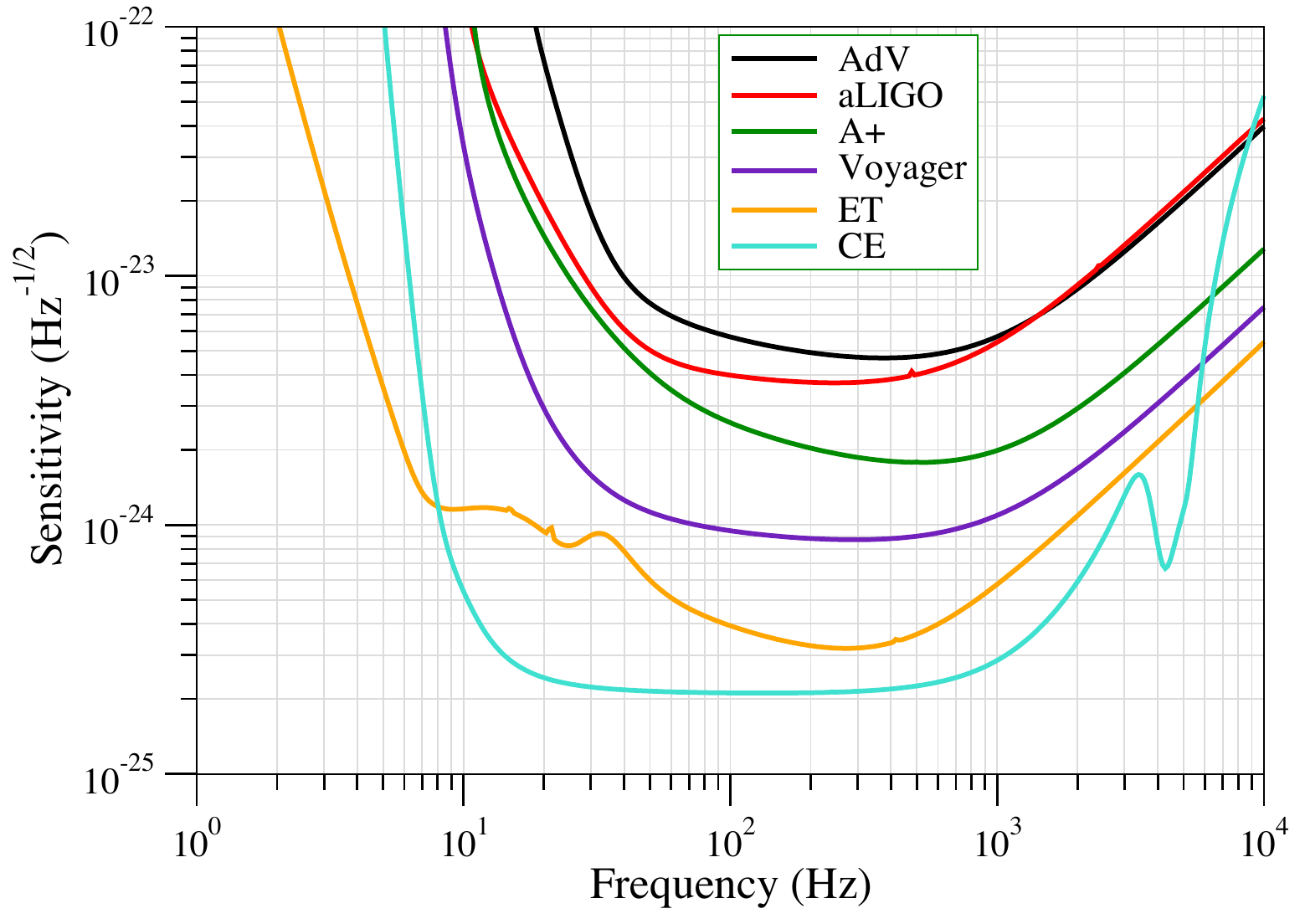}
    \caption{Strain sensitivity of three generations of ground-based gravitational wave detectors: (i) Advanced Virgo (AdV), Advanced LIGO (aLIGO) and A+, (ii) Voyager, and (iii) Einstein Telescope (ET) and Cosmic Explorer (CE). In the case of ET the sensitivity shown is that of an L-shaped detector with 10 km arms. The three V-shaped arms make the effective strain sensitivity a factor $3/2$ better (and the noise floor lower by the same factor).}
    \label{fig:sensitivities}
\end{figure}

Further upgrades to LIGO beyond A+ have been studied and they involve the development of new technology to mitigate thermal noise and gravity gradient background. Voyager \cite{Adhikari:2019zpy} is one such concept that could lead to a further increase in the sensitivity by a factor of $\sim 2$--5 over the frequency range 10 Hz to a few kHz (see Fig.\,\ref{fig:sensitivities}. At the moment we are not aware of any plans to upgrade Virgo or KAGRA and hence we will consider a network of five detectors: the three LIGO detectors operating with Voyager technology and Virgo and KAGRA in A+ mode. We will refer to this as the Voyager network, which will have access to several loud binary merger events. The Voyager network could constrain neutron star radius to within about 5\% or roughly 500 m for neutron stars between 1.5$M_\odot$ and 2.0$M_\odot$, as seen in Figure \ref{fig:combraderr}.

Improvements in sensitivity beyond the level of Voyager would require, among other technologies, longer arms and/or underground facilities, neither of which would be possible with the infrastructure that exists at the location of current detectors. The boldest of the new concepts are the Einstein Telescope (ET) in Europe and Cosmic Explorer (CE) in the US and, possibly, Australia. ET is an underground facility hosting three V-shaped detectors at the vertices of an equilateral triangle of 10 km sides \cite{Maggiore:2019uih}, while CE is a over-ground, L-shaped detector with 40 km arms \cite{Reitze:2020}. ET and CE will be roughly 10 to 30 times more sensitive than advanced detectors (cf.\,Fig.\,\ref{fig:sensitivities}) with the capability to observe hundreds of thousands BNSs mergers each year, many with SNRs larger than 100.

\subsection{Detector Networks}
Advanced LIGO, Advanced Virgo and KAGRA (LVK) have been taking data, albeit intermittently, since 2015, 2017 and 2019, respectively. They are expected to operate at design sensitivity during 2023-2024.  We have not included the measurement capability of this network as the number of loud (i.e., SNRs in excess of 25) BNS coalescences expected to be detected during the next science run (O4) is only $\sim $ few. 

%
LIGO-India, currently under construction, could join the upgraded A+ versions of the LVK network in the latter half of this decade; we shall call this the HLVKI+ network. Both LIGO and Virgo are planning for a further upgrade beyond 2030, referred to as Voyager in the US. A network in which Virgo and KAGRA operate at A+ sensitivities and LIGO-Hanford, LIGO-Livingston and LIGO-India operate at Voyager sensitivity, will be called VK+HLIv. This network will have the same performance as the one in which any three of 5 detectors are upgraded to Voyager and the remaining two operate at A+ sensitivity and we do not consider them separately. 

\begin{table}
    \caption{Upgraded and future gravitational-wave detectors whose ability to measure the EOS of matter in neutron star cores is evaluated in this study. The time-scale of operation of the various networks is our best guess estimate of when a given network is likely to operate; they do not correspond to any official projections.}
    \centering
    \begin{tabular}{l|c}
    \hline
    \hline
    \bf Detectors & \bf Network Name \\
    \hline
    LIGO (HLI+), Virgo+, KAGRA+      & \aplus\  \\
    LIGO (HLI-Voy), Virgo+, KAGRA+   & \voy\    \\
    ET, LIGO (HLI+), KAGRA+          &  \et\    \\
    CE, Virgo+, KAGRA+ LIGO-I+       & \ce\     \\
    ET, CE, KAGRA+, LIGO-I+          & \EC\     \\
    ET, CE, CE-South                 & \ECS\    \\
    \hline
    \end{tabular}
    \label{tab:networks}
\end{table}

Beyond 2035 one or more next generation observatories could begin to operate. To understand the relative merits of operating one or more such observatories we consider four different networks in which a subset of the current detectors operate at  A+ sensitivity at the same time as one CE (which we shall denote VKI+C), one ET (denoted HLKI+E), one each of CE and ET (denoted KI+EC) and a network consisting of one ET, one CE in the US and one CE in Australia (denoted ECS) without any A+ detectors.  In all, we consider six networks as enumerated in Table \ref{tab:networks}. For the ET and CE, we use fiducial locations and orientations as given in Ref. \cite{Borhanian:2020ypi}. We will next discuss the expected performance of various detector networks in detecting signals from and measuring the parameters of BNSs.

\subsection{Network Efficiency}
Gravitational wave detectors have a wide field of view of the sky but they are not equally sensitive to all directions. An interferometric detector like LIGO has a quadrupole antenna pattern and is able to detect only a fraction of all the sources from within a given distance. A network of non-collocated detectors increases the sky coverage and the five-detector network of HLVKI+ has an almost isotropic response. 

The efficiency of a detector network is a function of the luminosity distance (or redshift) and is defined as the fraction of all sources within a certain luminosity distance that can be (confidently) detected by the network, say with an SNR above a threshold SNR.  In order to compute the efficiency of a detector network we simulate BNS events with their parameters distributed as described in Sec.\,\ref{sec:waveforms}. The network SNR of an event is simply the quadrature sum of the SNRs in each detector:
\begin{equation}
\rho^2 = \sum_{A=1}^{n_D} \rho_A^2,\quad \rho_A^2 = 4 \int \frac{|\tilde h^A(f)|^2}{S^{A}_h(f)}\,df,
\end{equation}
where $\tilde h^A(f)$ is the Fourier transform of the response of detector $A$ to an incident gravitational wave [cf.\,Eq.\,(\ref{eq:response})], $S^{(A)}_h(f)$ is the one-sided noise power spectral density of detector $A$ as in Fig.\,\ref{fig:sensitivities}, $\rho_A$ is the matched filter SNR of the signal in detector $A$, $n_D$ is the number of detectors in the network, and $\rho$ is the network SNR. The efficiency of a detector is then defined as:
\begin{equation}
\epsilon(z) = \frac{1}{N} \sum_k \Pi(\rho_k(z)-\rho_T),
\end{equation}
where $N$ is the total number of simulated events, $\rho_k(z)$ is the network SNR for the k$^{\rm th}$ event, $\rho_T$ is the SNR threshold and  $\Pi$ is the step function, $\Pi(x)=0,$ if $x<0$ and $\Pi(x)=1,$ if $x>0.$ The SNR of an event depends not just on the redshift but on all other parameters of the source. In computing the network efficiency, we bin the SNR by redshift and ignore its dependence on all other parameters.  The SNR threshold $\rho_T$ serves as a proxy for detection confidence, larger SNRs are generally detected with greater confidence. We choose the threshold to be $\rho_T=12$---the minimum SNR required for a network of detectors to make a confident detection. While the SNR of 12 used here is required for a confident detection, it is not necessarily the SNR at which we can make the accurate measurements of tidal deformability necessary to determine a neutron star's radius and its EOS. In later sections, we will choose the best subset of all events to evaluate how well a network is able to measure the radii of neutron stars.

The efficiency of a network then also determines its detection rate. Within a given redshift, a network does not observe all the possible sources, but only a fraction $D_R$ given by:
\begin{equation}
D_R = \int_0^z \frac{r_z(z')}{(1+z')}\, \frac{dV}{dz'}\, \epsilon(z')\, dz'.
\label{eq:detection rate}
\end{equation}
We call $D_R$ the \emph{detection rate} of a network and it is  essentially the same as Eq.\,(\ref{eq:merger rate}) except that the integrand is weighted with the efficiency of the network.

Table \ref{tab:num-events} lists the number of events detected over a period of \textbf{two} years, as a function of detection threshold. An SNR of 12 is required for a confident detection, and at that level, the A+ network would observe about 800 sources over two years while the Voyager network would observe almost ten times as many. Meanwhile, a network containing at least one XG detector would observe about half all the sources within $z=1,$ (70,000 if XG is ET and 100,000 if XG is CE) (see Table \ref{tab:num-events}), and a network containing one ET and one CE would observe 30\% more sources than that. The ECS network would additionally observe about 10\% more sources than a network containing two XG detectors and 50\% more than a network containing a single XG detector.


\section{BNS Measurement Capability of Future Detector Networks} \label{sec:measurement}
In this Section we assess the measurement capabilities of different networks of gravitational-wave detectors introduced in Sec.\,\ref{sec:future}. We begin with a brief discussion of the distribution of the SNR in various detector networks followed by the accuracy with which parameters can be measured, in particular the effective tidal deformability. 

In the rest of the paper, we will only consider sources up to a redshift of $z=1.$ Within this redshift, we expect about 150,000 BNS mergers over a two-year period but the current rate uncertainty means this number could be 50\% larger or 25\% smaller. This is a redshift that is far greater than the horizon distance of A+ and Voyager networks while a network containing one or more of XG detectors would observe a vast majority of mergers within it. However, only a small fraction of them will have large enough SNRs to be useful for measuring the EOS.

\subsection{Signal to Noise Ratio Distribution for Nearby BNS Mergers}
\begin{figure*}
    \centering
    \includegraphics[width=2.\columnwidth]{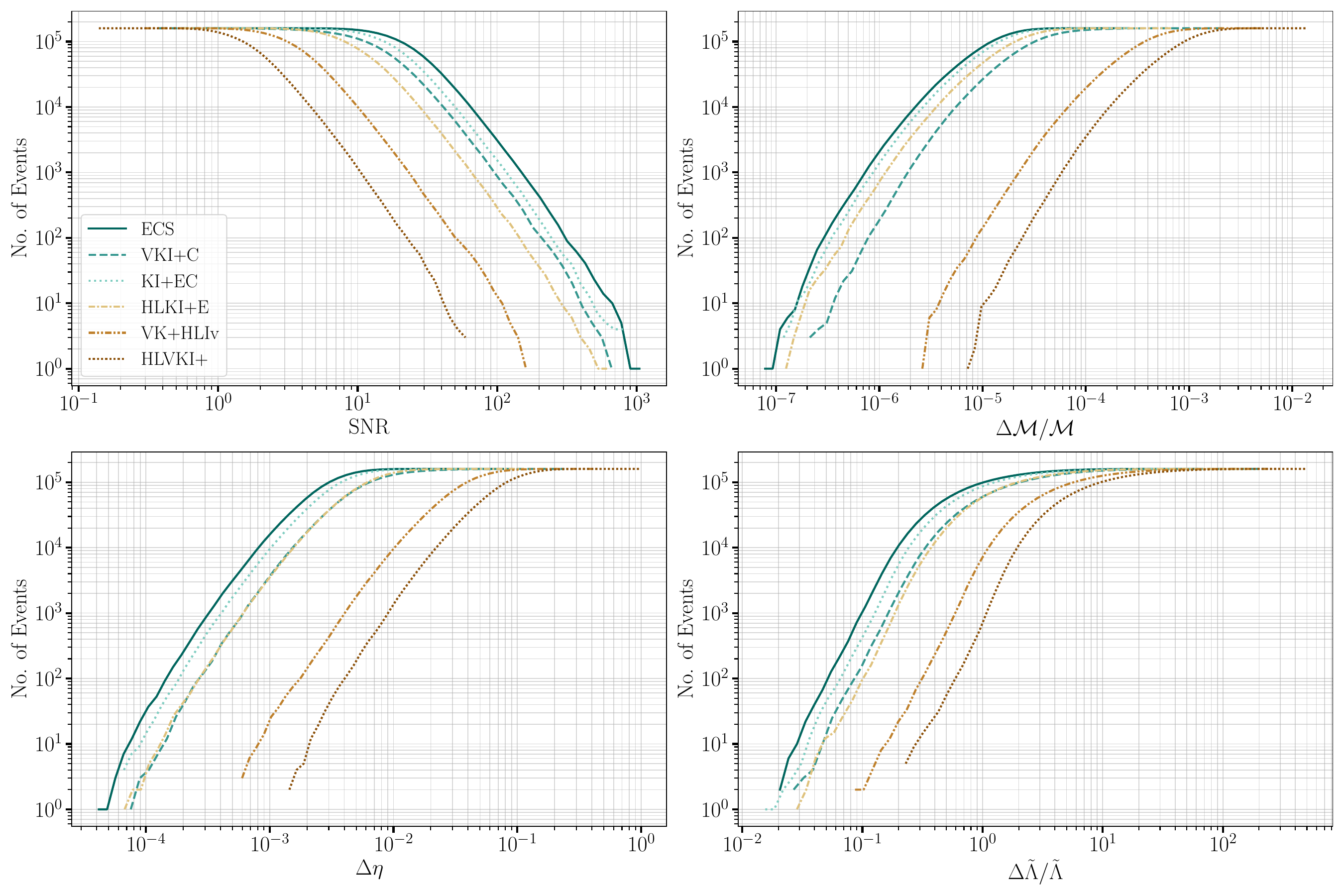}
    \caption{This plot shows the distribution of the measurement accuracy of the chirp mass $\cal M$, combined tidal deformability $\tilde\Lambda,$ symmetric mass ratio $\eta,$ and the SNR for 160\,000, events expected over a two year period, up to a redshift of $z=1.$ The source parameters are distributed as described in Sec.\,\ref{sec:waveforms}.}
    \label{fig:cdfs}
\end{figure*}

Figure \ref{fig:cdfs} plots the cumulative distribution of the SNR for the population of BNS mergers up to a redshift of $z=1$.  The VKI+C network should observe 10\% of the events with SNRs greater than 30 and 1\% of the events with SNRs greater than 60. In contrast, in the A+ network less than 0.1\% of events will have SNRs greater than 10. Cosmic Explorer and its southern counterpart operating along with Einstein Telescope would observe thousands of events each two years with SNRs greater than 100.

One must multiply the expected number of mergers within this redshift with the corresponding value of the CDF to get the number of sources expected to be observed each year. An estimate of actual number of events along with their SNR distribution is also given in Table ~\ref{tab:num-events}.

\begin{table}[bth!]
    \centering
    \begin{tabular}{r|r|r|r|r|r|r}
\hline
\hline
     $\rho_T$   &  \aplus & \voy & \et & \ce & \EC & \ECS \\
\hline
       12   & 840   & 7400 & 67,000 &  100,000  & 130,000  & 146,000  \\
       30   &  50   &  600 & 10,000  &  25,000  &  40,000  &  65,000  \\
       50   &   10   &   100 &  2,500  &   8000  &   12,000  &  23,000  \\
       100  &   0   &    10 &  300  &   1000   &    1,800  &   3800  \\
       300  &   0   &    0 &    10  &     50  &     70  &     150  \\
       500  &   0   &    0 &    1  &      5  &    10    &     30 \\
\hline
\hline
    \end{tabular}
    \caption{We list the number of events expected to be detected as we increase the SNR of events. Even with one Cosmic Explorer and/or Einstein Telescope, the number of BNS detections increases by an order of magnitude. In the bulk of this work, we focus our analysis on top 100 events with the highest SNR for each detector network. This cut corresponds to SNR of 100 or more for networks with at least on XG-era detector and about 50 or below for A+ detectors.}
    \label{tab:num-events}
\end{table}

\subsection{Fisher Information Approach for Measurement Accuracy}

Our goal is to estimate the accuracy with which 
parameters of an event can be measured by gravitational-wave detector networks. To this end, we employ the Fisher information matrix approach \cite{Arun:2004hn}, which allows a reliable estimation of errors when the SNRs large (say more than about 30 or 50). We use the open source software \textsc{gwbench} \citep{Borhanian:2020ypi} to generate and sample posteriors for a set of randomly selected signals. \textsc{gwbench} is a software package that computes the Fisher information matrix (FIM) $\mathcal F$ whose inverse gives the variance-covariance matrix. The starting point of the computation is the response of a detector to incident gravitational wave with polarizations $h_+$ and $h_\times$:
\begin{equation}
h^A(t,{\bm\theta}) = F_+^A(t,\alpha,\delta,\psi) h_+(t,{\bm\mu}) +  F_\times(t,\alpha,\delta,\psi) h_\times(t,{\bm\mu}) 
\label{eq:response}
\end{equation}
where $A$ is an index denoting the detector in question.  Here $F_{+,\times}$ are the plus and cross antenna pattern functions of the detector that depend on the right ascension $\alpha$ and declination $\delta$ of the source, and the polarization angle $\psi.$ The time dependence of the antenna pattern functions are only important when the motion of the detector relative to the source is perceptible, and for sources that last for more than 30 minutes. The polarization amplitudes $h_+$ and $h_\times$ depend on the intrinsic parameters of the sources such as the masses $m_1$ and $m_2$ of the companion stars, \footnote{In principle the companions can have spin angular momenta, but neutron stars are not expected to have large spins and they are not included in this study.} and the effective tidal deformability $\tilde\Lambda,$ but also the extrinsic parameters that include the orientation $\iota$ of the binary's orbit relative to the line-of-sight from the Earth to the source and the source's luminosity distance $D_L.$ These are all combined in the parameter ${\bm\mu}=\{$$\cal{M}$$, \eta, \tilde\Lambda, \iota, D_L\},$ where instead of the companion masses we have used the symmetric mass ratio $\eta \equiv m_1 m_2/M^2,$ and the chirp mass $\cal M$$ \equiv \nu^{3/5}M$ ($M\equiv m_1+m_2$).  The parameter set ${\bm\theta}$ captures all the parameters describing the response of a detector to an incoming gravitational wave (see below for the full list of parameters). 

Given the Fourier domain representation $h^A(f; \Vec{\theta})$ of the detector response, the Fisher matrix is given by:
\begin{equation}
    \mathcal{F}^A_{ij} = \left<\frac{\partial h^A(f)}{\partial \theta^i}, \frac{\partial h^A(f)}{\partial \theta^j} \right>\,, 
\end{equation}
where the inner product of any two functions $a(f)$ and $b(f)$ is defined as
\begin{equation}
    \left<a(f),b(f)\right> = 2\int_{f_{\rm low}}^{f_{\rm high}} \frac{a(f)^*b(f)+a(f)b(f)^*}{S^A_h(f)} df.
\end{equation}
where $a^*(f)$ denotes the complex conjugate of $a(f).$
The Fisher matrix of a network of detectors is simply the sum of the matrices corresponding to individual observatories  in the network, i.e.
\begin{equation}
{\cal F}_{ij} = \sum_A {\cal F}_{ij}.
\end{equation}

Given the Fisher matrix, the covariance matrix $\mathcal{C}_{ij}$ among the parameters is the inverse of the Fisher matrix, i.e. $\mathcal{C}_{ij}=\mathcal{F}_{ij}^{-1}.$ 

To construct the Fisher likelihood surface, we choose a low-frequency cutoff, $f_{\rm low}$, of $10~\rm Hz$ for A+ and Voyager detectors and $5~\rm Hz$ for XG detectors. The high-frequency limit is taken to be the maximum allowed frequency given the sampling rate (typically chosen to be 4096 Hz), but the signal model never extends to such high frequencies even for the lowest-mass neutron stars considered in this paper. We then compute a 10-dimensional Fisher likelihood consisting of the parameter set ${\bm\theta}=\{$$\cal M$$, \eta, \tilde{\Lambda}, D_L, \psi, \cos\iota, \alpha, \delta, \phi_c, t_c\},$  where  $t_c$, and $\phi_c$ are the fiducial  time of coalescence, and the gravitational-wave phase at coalescence, respectively.

\subsection{Measurement Accuracy of Simulated Population}
Fig.~\ref{fig:cdfs} plots the errors on the parameters of the simulated population in the form of distribution functions. We have shown the results for a subset of all the parameters that are relevant to the measurement of the mass-radius curves. These are the chirp mass $\cal M$, the symmetric mass ratio $\eta$ and the effective tidal deformability $\tilde\Lambda$. We see a clear delineation in the measurement capabilities of current and upgraded networks and XG observatories. The precise measurement of the parameters is, of course, accomplished by tracking the phase evolution of the binary. The chirp mass and mass ratio are most accurately measured if the number of cycles in the band is large (i.e. if the signal's phase can be tracked over longer periods) and a good improvement in low-frequency sensitivity for XG detectors is responsible for this vast improvement in the measurement of the mass parameters.  The reduced tidal deformability measurement comes from the signal's phase evolution close to merger, or the high-frequency part of the signal, which will be clearly visible in XG detectors. 

The remaining parameters---sky position, distance, and orientation of the binary in the plane of the sky---also show a clear delineation between detector generations, except the instance where the addition of CE without ET performs similar to Voyager networks\footnote{The performance equivalence argued here is for a fraction relative to the total number of detected events. In absolute terms, even a single CE will have outstandingly more events with a given measurement error.}. 

\paragraph{Sky localization} For very short transient signals, the sky localization is measured using the gravitational-wave travel times between different detectors and, therefore, depends on the number of non-collocated detectors. Thus, the 5-detector network of VK+HLIv, achieves greater precision than a 4-detector XG network VKI+C, although the signal strengths in the latter are much greater. For longer signals that make a discernible trail on the sky, the variation of the antenna response across the sky can be used to improve the sky position of the source. Since ET is more sensitive between 5~Hz and 8~Hz, where a typical BNS signal ($1.4~\rm M_{\odot} + 1.4~\rm M_{\odot}$) spends more than an hour ($\sim75$ minutes), a trail spanning more than $15^{\circ}$ on the sky (or, a fifth of the total variation in the antenna pattern) is clearer in the presence of an ET detector. Moreover, HLKI+E is composed of five detectors, which accentuates the sky resolution.

\paragraph{Inclination angle} The measurement of the inclination angle is dependent on the distinguishability of the two gravitational-wave polarizations. Since ET is a triangular detector that measures three independent strains, each strain has different polarization content, leading to an accurate estimate of the polarization content and, thereby, the inclination angle. A CE detector alone cannot distinguish between the two gravitational-wave polarizations and it is the 2G background (inclined with respect to each other and CE) that provides crucial assistance to the VKI+C network in the polarization measurement. However, a mutually inclined 5-detector network VK+HLIv still achieves greater precision than a 4-detector VKI+C network.

\paragraph{Luminosity distance} The luminosity distance parameter is most correlated with the inclination angle. Hence, a precise measurement of the inclination angle also leads to an accurate measurement of the luminosity distance. Thus, the measurement trends for the luminosity distance across networks follows the trends in the inclination angle.

\section{Inferring Neutron Star EOS from Mass-Radius Curves}
\label{sec:tidal deformability and radius}

The Bayesian inference of the chirp mass $\cal M$, and symmetric mass ratio $\eta$ of the BNS events detected by LIGO and Virgo are the most precise measurements among all parameters of BNS events. While the effective tidal deformability is not measured as precisely, upcoming gravitational-wave detector networks promise vastly improved measurements (cf. Fig.\ref{fig:cdfs}). To measure the radii of component stars, however, it is necessary to know what the individual tidal deformabilities $\Lambda_1$ and $\Lambda_2$ are as well as the tidal Lover number $k_2$ (cf.~Eq.~ \ref{eq:lambdatilde}). Unfortunately, gravitational-wave observations can only provide a reliable estimation of the linear combination $\tilde \Lambda.$ This problem has been resolved temporarily via the proposal of a set of quasi-universal relations for neutron stars, which are approximately obeyed by hundreds of current models of the EOS \cite{Yagi:2016bkt}.

In this set, there are basically two universal relations. The first of these relates the asymmetric combination  of the individual tidal deformabilities\footnote{We follow the convention $m_1 > m_2$ and, consequently, $\Lambda_1 < \Lambda_2.$}  $\Lambda_a\equiv (\Lambda_2-\Lambda_1)/2$ to the symmetric combination $\Lambda_s\equiv (\Lambda_2+\Lambda_1)/2$ via the mass ratio $q$
\begin{equation} 
\label{eqn:unirel1}
    {\Lambda}_a = F_{{n}}(q){\Lambda_s}\frac{a+\sum_{i=1}^{3}\sum_{j=1}^{2}b_{ij}q^j{{\Lambda}_s}^{-i/5}}{1+\sum_{i=1}^{3}\sum_{j=1}^{2}c_{ij}q^j{\Lambda_s}^{-i/5}},  
    \end{equation}
where the function $F_n(q)$ is given by
\begin{equation} 
    F_n(q) = \frac{1-q^{10/(3-n)}}{1+q^{10/(3-n)}}.
\end{equation}
The fitting parameters $b_{ij}, c_{ij}, a$ and $n$ are given in Table I of Ref.~\cite{Chatziioannou:2018vzf}.
The second universal relation \cite{Maselli:2013mva} relates the compactness $C\equiv GM/(c^2R)$ of an individual neutron star to its tidal deformability:
\begin{equation} \label{eqn:unirel2}
C(\Lambda)= \sum_{k=0}^{2} a_k (\ln{\Lambda} )^k,  
\end{equation}
where the fitting parameter $a_k$ are also given in Table I of Ref.~\cite{Chatziioannou:2018vzf} (also see Ref.~\cite{Pradhan:2022rxs} for similar relationships).

The first of the universal relations Eq.~(\ref{eqn:unirel1}) can be used to decouple the effective tidal deformability into individual tidal deformabilities. Then the second universal relation Eq.~(\ref{eqn:unirel2}) can be used to compute the radius. These universal relations, however, have been shown to introduce systematic errors \cite{Kastaun:2019bxo} that must be corrected in order to obtain an unbiased estimation of the EOS \citep{Kashyap:2022wzr}. In the rest of this section, we describe our simulation method to assess the radii measurements for a set of future gravitational-wave observatories with corrections for these errors.

\subsection{From Gravitational Wave Measurements to Neutron Star Radii}

We begin with the Fisher information matrices (FIM), computed using the \textsc{gwbench} software, for the entire simulated BNS population and all the detector networks described in Sec.\,\ref{sec:injections} for  a set of three EOS models and the \textsc{IMRPhenomPv2NRTidal} waveform model.  Diagonal elements of the covariance matrix (inverse of the FIM) are the standard deviations of the source parameters: $($$\cal M$$, \eta, \tilde{\Lambda}, \phi_c, t_c, D_L, \cos{\iota},\alpha, \delta, \psi)$. In order to obtain radii of the companion stars from the parameters measured via gravitational-wave observation, we simulate posterior samples by generating a multi-dimensional Gaussian sample using the injection values as mean values and the inverse of the FIM as the covariance matrix. We need only three of these parameters ($\cal M$$, \eta, \tilde{\Lambda}$) for the estimation of radii. To break the degeneracy between two tidal deformabilities and get individual radii, we follow the procedure described in \cite{Kashyap:2022wzr} (see also \cite{LIGOScientific:2018cki} for an alternative method), which is briefly described below. 

First, in the expression for \lamt we eliminate $\Lambda_1$ and $\Lambda_2$ in terms $\Lambda_s$ and $\Lambda_a.$ We then use the first universal relation in Eq.~(\ref{eqn:unirel1}) to replace $\Lambda_a$ with $\Lambda_s$ in the expression for \lamt, thereby writing \lamt as a function of only $\Lambda_s$ and $q$. Since gravitational-wave observations measure \lamt, we can invert the expression for $\tilde\Lambda=\tilde\Lambda(\Lambda_s, q)$ to get $\Lambda_s(\tilde{\Lambda},q).$ Thus, from gravitational-wave measurements of the mass ratio and the effective tidal deformability we can extract the symmetric combination $\Lambda_s$ and then, using Eq.~(\ref{eqn:unirel1}), also $\Lambda_a$. These two are then inverted to obtain the individual tidal deformabilities of the component stars. Thereafter, we use the $C$-$\Lambda$ universal relation in Eq.~(\ref{eqn:unirel2}) to derive the compactness and, with the individual masses, obtain the posterior probability distribution of the radii for component neutron stars.

\subsection{Correcting Systematic Errors in Neutron Star Radii}
\label{sec:inferring the EOS}
Universal relations introduce systematic errors in the estimation of individual tidal deformabilities and radii which will dominate the source of errors in the era of XG observatories \citep{Kashyap:2022wzr}. Due to the fact that $\delta\tilde\Lambda$ cannot be measured accurately, it is not possible to obtain a truly, arbitrarily precise, model-agnostic measurement of neutron star radii or compactness using only gravitational-wave measurements\footnote{Note that even if $\Lambda_1$ and $\Lambda_2$ are measured by gravitational-wave observations the tidal Love numbers of the two neutron stars will still be unknown and hence the radii cannot be inferred}. However, it turns out that for the purpose of EOS model selection the systematic errors can be corrected as we will briefly argue below (see Ref.~\cite{Kashyap:2022wzr} for details). 

As discussed before, the \textsc{gwbench} framework is used to create a population of BNS events in which the tidal deformability $\Lambda$ of each neutron star of mass $m$ is computed for a specific EOS model (one of ALF2, APR3 or APR4).
Out of the 150,000 simulated events, we choose 100 events that have either the greatest SNR or the best-measured tidal deformability. For the 100 events, we will have 200 mass-radius posteriors, one for each of the companion stars. 
We then sample a discretized mass-radius curve containing 200 points by randomly sampling each star’s mass-radius curve and repeat the process to generate a large number of realizations, representing the mass-radius curve supported by the 100 chosen events.
Sampling in this manner can form mass-radius curves which violate causality, and thermodynamic constraints.
However, we note that this makes our estimates more conservative, and curves which differ greatly from the true EOS as a result of this will be rejected by the chi-square statistic described in the following section.
The radii used to construct these mass-radius curves then contain the systematic errors introduced by our use of the universal relations, so the resultant mass-radius curve will also be biased.
Given an EOS, we can determine the exact value of this bias by comparing the mass-radius curve for an EOS generated using the TOV equations to that of a curve generated using the universal relations.
With this in hand, we can calculate the correction necessary to account for the systematic errors introduced by the universal relations which, when applied to a mass-radius curve, will closely match the exact TOV curve.

In this work, we thus correct for these systematic errors by applying these corrections to the calculated mass-radius curves per EOS.
For example, if we would like to determine whether the underlying equation of state of our mass radius curve is ALF2, we first apply the known correction for ALF2 to our mass-radius curves and then complete the comparison described in the next section.
If the true underlying EOS is not the one for which we have applied the correction, then the correction will not correctly account for the systematic errors and we can only assume that most similar resulting mass-radius curve is the closest to the excluded true model.

We will consider the true model in turn to be one of the 10 EOS models shown in Fig.~\ref{fig:EOS_mr} and show how the corrected-mass-radius curves compare with the true EOS model.  In practice, one has to compare the curves with the full set (of millions) of curves. In order to clearly illustrate the power of the method, we have not done so and instead reserved a more detailed and careful Bayesian statistical analysis of model selection in an upcoming publication.

\subsection{EOS Model Selection Using \texorpdfstring{$\mathbf{\chi^2}$}~ Statistic}
After generating a mass radius curve as described in the previous section, we must compare it to a set of EOS models in order to determine the true EOS of the population.
We complete this comparison with the following statistic:
\begin{equation}\label{eqn:chisq}
\chi^2_{k,M} = \frac{1}{N} \sum_{i=1}^{N}  \frac{(r_{i}^{k}-r_{i}^{M})^2}{\sigma_{i}^{2}}
\end{equation}
Here, $N$ is the number of events, $k$ stands for one of the realizations constructed from the mass-radius posterior and $\sigma_i$ are 1$-\sigma$ uncertainty in the radii calculated after applying the systematic bias correction. We generate 500 realizations of the mass-radius curve and obtain a distribution of the $\chi^2$ statistic for each of the 10 EOS models. 

If a realization of the mass-radius curve is close to the model to which it is compared to, the numerator of Eq.~(\ref{eqn:chisq}) becomes zero. If, however, the uncertainties in the tidal deformability are large, the $\chi^2$ again becomes small regardless of the position of $m$-$\Lambda$ posterior distribution with respect to the model $m$-$\Lambda$ curve. This is a drawback in our model and leads to the underestimation of near-future LVK upgrades in distinguishing EOS models. Therefore, when comparing against a collection of EOS, the smallest $\chi^2$ value should correspond to the injected EOS for XG detector configurations in which statistical errors are much smaller and recovery of EOS in the data is more accurate, but for near-LVK upgrades, this may not be true due to the large errors in tidal deformability. We discuss our results in the next section and defer the improvement to the Bayesian formulation of our $\chi^2$ method to future studies. 

\subsection{Combining Results from Multiple Events}
\label{sec:comberr}

The accuracy of radii posteriors depends to a large extent on the accuracy of tidal deformability measurements, which in turn depends on mass-posteriors. Heavier component masses have smaller tidal deformabilities, which are difficult to measure. The low accuracy of the tidal deformabilities results in poorer radii measurements, which constrain the high-density regime of the EOS, while lighter component masses typically result in better measurement of the radii. The correct reflection of the radii uncertainty, therefore, cannot be at some fiducial mass but will be a function of the companion mass.  Having measured the radii of several hundred neutron stars, it is possible to get a better handle on the radius at a fixed mass. 

Evidence from the observation of multiple events, in principle, can be combined to give us integrated evidence of the constraints on neutron star radii. In this paper, we bin the selected set of events over the range of companion masses from 1.0 M$_\odot$ to the maximum mass supported by the EOS in steps of 0.05 M$_\odot$ wide bins and assume that all neutron stars in a given bin have the same radius. The uncertainty in the radius in each bin is computed as the quadratic harmonic sum of individual 1-$\sigma$ uncertainties in the radius of individual neutron stars that lie within the bin. This procedure is equivalent to combining the posteriors of radii corresponding to all the NSs in a particular mass bin assuming priors are the same for all NSs. While not ideal, this method improves upon the method used in \citep{De:2018uhw}, which assumes that radii of neutron stars over the entire mass range from 1.1 M$_\odot$ to 1.6 M$_\odot$ are the same. We note that this latter assumption could introduce an intrinsic systematic error of 200 m (Eq.-6 of \cite{De:2018uhw}) --- a value much larger than the measurement uncertainty we find in the case of XG detectors. We report the results of this calculation in the next section. The accuracy of radii measurements can be translated to the accuracy in the estimation of nuclear physics parameters \cite{Rose:2023uui, Sabatucci:2022qyi} which we defer for future work.

\begin{figure*}
    \includegraphics[width=0.99\textwidth]{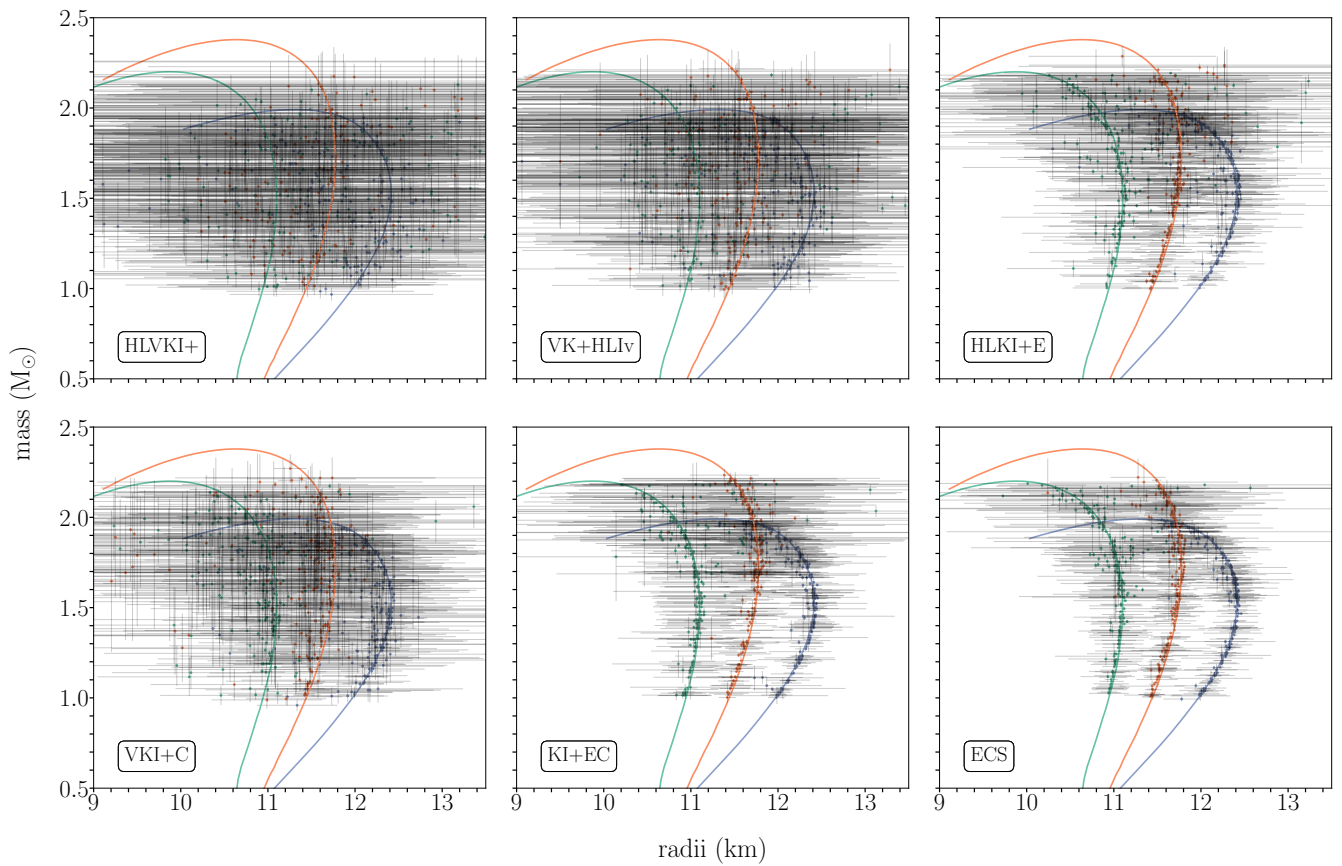}
    \caption{TOV mass-radius curves of ALF2(blue), APR3(orange), and APR4(green) overlaid with the bias-corrected recovered mass and radius as well as their errors (grey bars) in a subset of near-future and XG detector networks, for a set of 100 random events drawn from the 500 loudest SNR. There is a clear trend of improving radius error as the detector networks improve left to right, top to bottom. Additionally, in the best detector networks, radius errors also improve with decreasing mass, as is to be expected with higher accuracy in the measurement of higher tidal deformability.} 
    \label{fig:MR_error_loudest}
\end{figure*}

\subsection{Impacts of assumed cosmology}
\label{sec:cosmology}
To obtain the error in the radius measurement, we need to convert the uncertainties in the detector-frame chirp mass to that of the source-frame chirp mass. In doing so, we have assumed that the cosmological parameters, like the Hubble constant $(H_0)$, are known exactly (see Sec. IIA). Although advancements in gravitational-wave detector networks are expected to achieve sub-percent precision in measuring cosmological parameters \cite{Borhanian:2020vyr, Gupta:2023lga, Gupta:2022fwd, Branchesi:2023mws, Dhani:2022ulg, Muttoni:2023prw}, the associated uncertainties may still impact radius measurements.\\
Note that the two most precise measurements of $H_0$, from the Planck mission \cite{Planck:2018vyg} and the SH0ES project \cite{Riess:2021jrx}, are in disagreement at the $5-\sigma$ level, which is called the Hubble tension. To obtain a liberal estimate of how the uncertainty in $H_0$ can affect radius measurements, we perform Bayesian parameter estimation with \texttt{BILBY} \cite{Ashton:2018jfp, Romero-Shaw:2020owr} for a (1.45, 1.35) M$_\odot$ BNS system, at 400 Mpc, with APR4 as the assumed EOS. For this zero noise analysis, the system is injected in a network with one Einstein Telescope and two Cosmic Explorer observatories (SNR ~ 330). The injected system is made to obey the SH0ES estimate of $H_0 = 73.3\,\mbox{km\,s}^{-1}\,\mbox{Mpc}^{-1}$, whereas the recovery is performed assuming the Planck18 value of $H_0 = 67.4\,\mbox{km\,s}^{-1}\,\mbox{Mpc}^{-1}$, i.e., a fractional error in $H_0$ of $\sim 8\%$. Employing the same analysis as in the current study, we obtain the $68\%$-credible region for radius estimate to be ~370m ($\Delta R/R ~ 3\%$). In contrast, the bias in the estimate due to inference with the incorrect cosmology is ~60m. Thus, even at an exaggerated uncertainty of $8\%$ in $H_0$, we see that the statistical uncertainty in the radius measurement outweighs the resulting bias. Therefore, at the forecasted precision levels of cosmological parameter measurement with next-generation observatories, we do not expect the uncertainty in their estimation to play a significant role in the estimation of the radius of the neutron star.

\section{Results from a Population Study}
\label{sec:results}
In this section, we present the accuracy of radius measurements inferred from a sub-population of 100 best events, for six different detector networks and three different EOS models. The sub-population is chosen to be either events with the best-measured tidal deformabilities or the largest SNRs. In order to gauge Monte Carlo errors, we start with a set of 500 events satisfying the aforementioned criteria and then bootstrap several realizations of 100 events. We present the results in a series of plots that compare the measured mass-radius curves to those derived from different EOS models, the $\chi^2$ histogram between the measured and model radii, and precision with which radius can be measured by combining events in $0.05 M_\odot$-wide mass bins for different EOS models.

\begin{figure*}
    \includegraphics[width=0.99\textwidth]{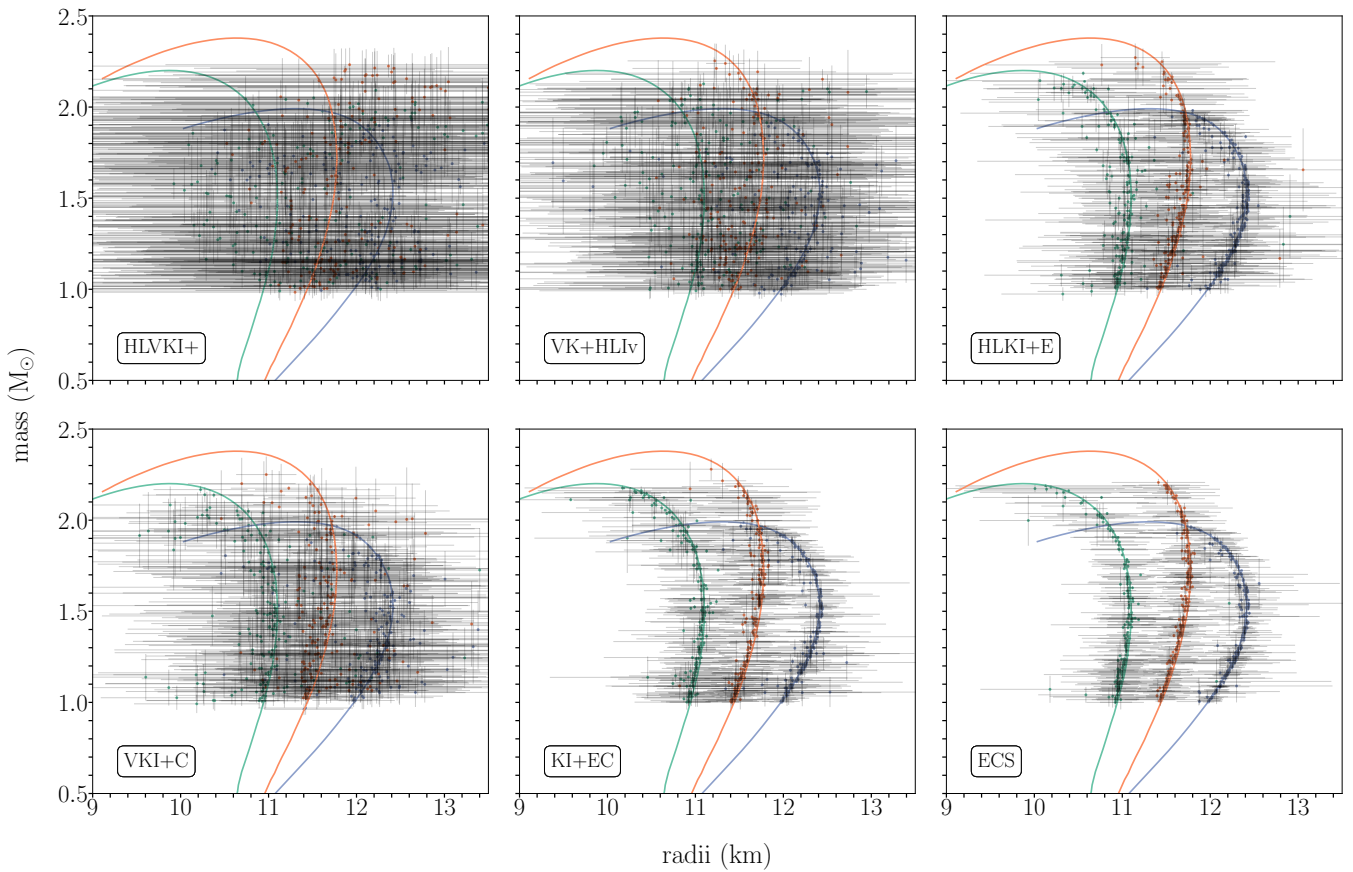}
    \caption{Same as Fig.\,\ref{fig:MR_error_loudest} except the 100 out of 500 events with best measured tidal deformability are chosen. Again, there is a clear trend of improving radius error as the detectors network improves left to right, top to bottom. Note that the trend of improved radius error with decreased mass is not clear here as it was with the loudest in the SNR set. This is a natural result from the selection of only the best measured combined tidal deformability systems as opposed to those with the best SNR as in Fig \ref{fig:MR_error_loudest}.}
    \label{fig:MR_error_bestmeasured}
\end{figure*}

\subsection{Radius Measurement}

Figures \ref{fig:MR_error_loudest} and \ref{fig:MR_error_bestmeasured} plot the uncertainties in the measurement of masses and bias-corrected radii for 100 random events drawn from the 500 events, with the largest SNR and the best-measured tidal deformability, respectively, for a population of BNS described in section \ref{sec:injections}. The cumulative distribution of the measurement uncertainties in the parameters used in this calculation are shown in Fig \ref{fig:cdfs}. Multiple realizations of the 100 events (out of 500) do not show significant differences in the mass-radius curves and hence we have shown the plots for just one realization. Results are shown for the six different detector networks. In each case, the true model is in turn chosen to be ALF2 (blue), APR3 (orange), or APR4 (green). In these plots, we show the bias-corrected radii using only the injected models as described in \ref{sec:inferring the EOS}. Otherwise, the plot would be too busy; the chi-square plots, to be discussed below, will compare the bias-corrections applied to radii assuming the true EOS model to be any one of the three candidates. Measurement uncertainties in mass and radius are plotted in grey. 

Figures  \ref{fig:radius_chirpmass_eta_snr_bestLambdaTilde}  and  \ref{fig:radius_chirpmass_eta_snr_loudest}  in Appendix \ref{sec:appendix}  show the same result but plotted in the chirp mass-symmetric mass ratio space, while  Figures  \ref{fig:radius_Lambda_chirpmass_snr_bestLambdaTilde}  and  \ref{fig:radius_Lambda_chirpmass_snr_loudest}  show the results in the chirp mass-combined tidal deformability space.  Figures \ref{fig:radius_chirpmass_eta_snr_bestLambdaTilde} and \ref{fig:radius_Lambda_chirpmass_snr_bestLambdaTilde}  are for events with the best-measured tidal deformability while Figures \ref{fig:radius_chirpmass_eta_snr_loudest}  and \ref{fig:radius_Lambda_chirpmass_snr_loudest}  are for events with the largest SNRs. The color shade of the dots in these plots represents the radius uncertainties while the size of the dots is a measure of the SNR of the events as shown in the legend. 

Note that these results are based on the Fisher Matrix calculation of the measurement uncertainty. Therefore, the results we see here can be taken as a lower bound of what we might actually expect from a full Bayesian analysis of parameter estimation of these events.

From Figs. \ref{fig:MR_error_loudest} and \ref{fig:MR_error_bestmeasured}, there is an evident trend of marked improvement in the measurement of the radii as the detector networks themselves improve. The recovered radii fall closer to the injected EOS curve, and the measurement uncertainties vastly decrease as the number of XG observatories in a network rises from 0 to 3. Notably, the maximum uncertainty in the radii, most easily read from the color bars of Figures \ref{fig:radius_chirpmass_eta_snr_bestLambdaTilde}-\ref{fig:radius_Lambda_chirpmass_snr_loudest}, vary from, in the worst detector, about 2500~m to, in the best network, only about 300~m. At low masses, the disparity is especially clear, and this is a natural result for these  networks--- particularly the improvement once at least one XG detector added to the network.

It is notable that in networks which contain just one XG detector, the HLKI+E network slightly outperforms that of VKI+C in the measurement of radius error. This is expected for two reasons. First, the HLKI+E network contains one additional detector than that of VKI+C, which inherently improves its sensitivity. Second, the ET sensitivity curve, as seen in Fig. \ref{fig:sensitivities}, contains a long tail in the low-frequency regime not present in the CE curve. This increases the time neutron star signals spend in the band, and results in a better-measured chirp mass and, therefore, better-measured radii. The evidence of this can be seen in the chirp mass and radii CDFs of Figure \ref{fig:cdfs}. There, the HLKI+E chirp mass CDF shows clearly a smaller relative error than that of VKI+C, and where the HLKI+E tidal deformability CDF shows on level or slightly smaller relative error than that of VKI+C.

In the data set with the loudest SNR events (Figures \ref{fig:MR_error_loudest}, \ref{fig:radius_chirpmass_eta_snr_loudest}, and \ref{fig:radius_Lambda_chirpmass_snr_loudest}), higher-mass systems are less constrained---especially in radius---than lower-mass systems, while this is not necessarily true for the set of best-measured tidal deformability events (Figures \ref{fig:MR_error_bestmeasured}, \ref{fig:radius_chirpmass_eta_snr_bestLambdaTilde}, \ref{fig:radius_Lambda_chirpmass_snr_bestLambdaTilde}). Again, this is an expected result, as we accumulate most SNR for BNS systems during the low-frequency inspiral phase, while the best measurements of tidal deformability come from the high-frequency part of the waveform during the merger. Thus, a high SNR does not beget a well-measured tidal deformability or radius. Additionally, although gravitational-wave amplitudes for high-mass systems tend to be larger compared to low-mass ones, the value of their tidal deformability tends to be smaller. These small values combined with short inspiral times result in larger relative errors in the measurement of tidal deformability and radii despite the boost in SNR from higher amplitudes. This trend is especially clear in Figs. \ref{fig:radius_Lambda_chirpmass_snr_bestLambdaTilde} and \ref{fig:radius_Lambda_chirpmass_snr_loudest} where in Fig. \ref{fig:radius_Lambda_chirpmass_snr_loudest} the highest radii errors for each panel (in yellow) are always seen in the right, or the high mass and low tidal deformability, portion of the plot while in Fig. \ref{fig:radius_Lambda_chirpmass_snr_bestLambdaTilde} the worst measured events (again in yellow) are spread throughout parameter space.

Similarly, in Figs. \ref{fig:radius_chirpmass_eta_snr_bestLambdaTilde} and \ref{fig:radius_chirpmass_eta_snr_loudest}, it appears that a high symmetric mass ratio, and high chirp masses may result in poorly measured tidal deformability for the highest SNR events, but not necessarily for those with the best measured tidal deformability. In Figure \ref{fig:radius_chirpmass_eta_snr_loudest}, large radii errors (in yellow) are typically grouped in the upper right-hand corner of most plots, with a small spread along the right-hand edge in the ALF2 and APR4 EOS, and a small line along the upper-edge in the VKI+C network of ALF2. This is again due to the previously discussed issue with taking the loudest SNR events, but whether this is individually caused by either the high symmetric mass ratio or the large chirp masses is not immediately clear. As previously mentioned, a high chirp mass comes with a small tidal deformability and therefore large relative error. However, a high symmetric mass ratio can also decrease the inspiral time, or time in a frequency band, and therefore again the accuracy of the measurements becomes low. Notably, in the set of best measured tidal deformability shown in  Fig.~\ref{fig:radius_chirpmass_eta_snr_bestLambdaTilde}, the large errors are distributed more evenly throughout the plot and have lower maximums than their high SNR equivalent. 
 

\begin{figure*}
\includegraphics[width=0.90\textwidth]{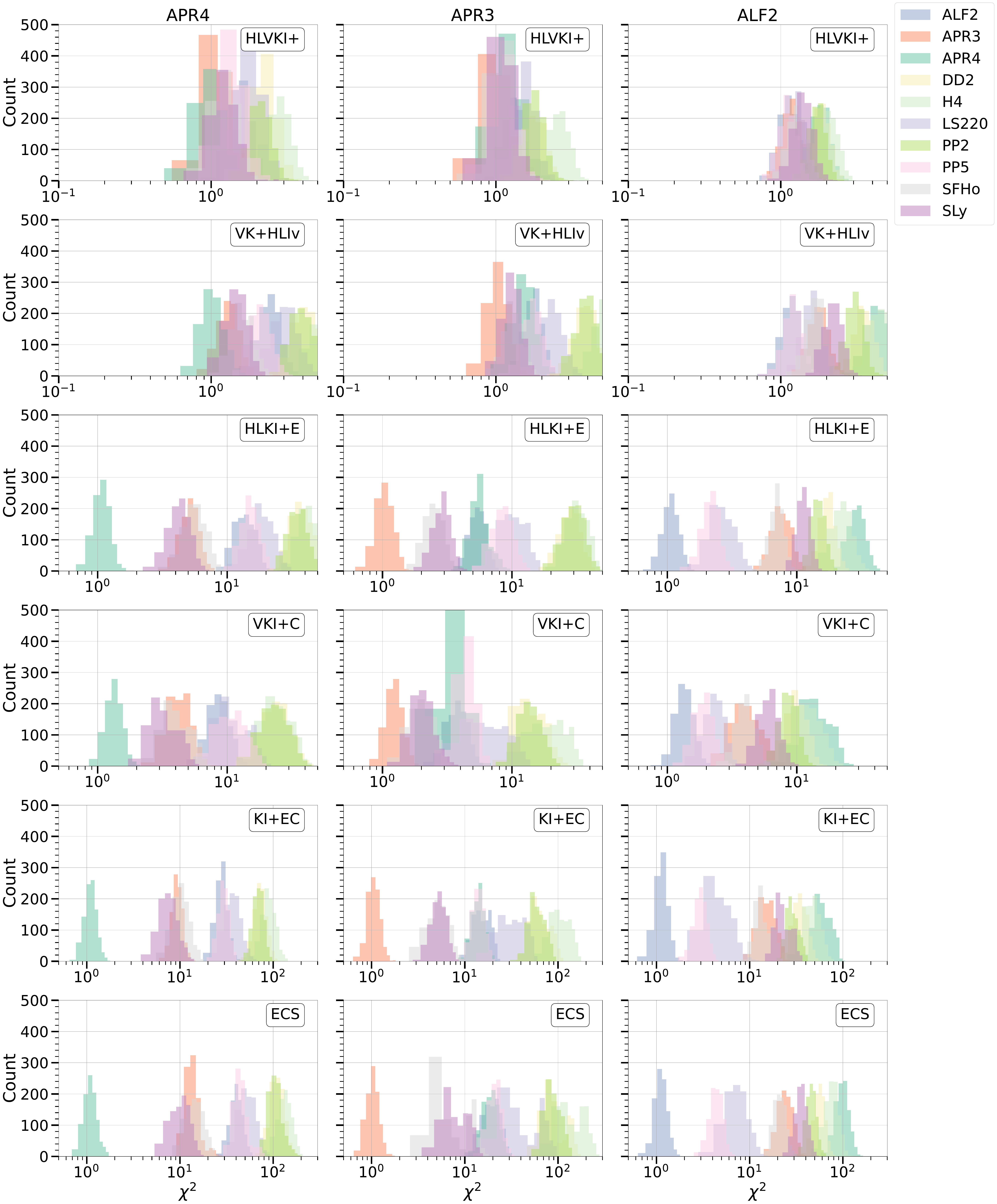}
\caption{Chi-square histograms for 100 events from those 500 with the smallest error in combined tidal deformability. The injected EOS is listed along the top, and the colored histograms represent the result assuming a second EOS model, including the original injection. Detector networks are organized by sensitivity row-wise with the most sensitive network at the bottom. In every EOS and network scenario including at least one XG detector, the injected EOS is recovered correctly and easily distinguishable from the other nine via this test. In our two least sensitive and nearest future detector networks, HLVKI+ and VK+HLIvc, the opposite is clearly true and all models are indistinguishable.}

\label{fig:chisq-lamt}
\end{figure*}

\begin{figure*}
\centering
\includegraphics[width=0.80\textwidth]{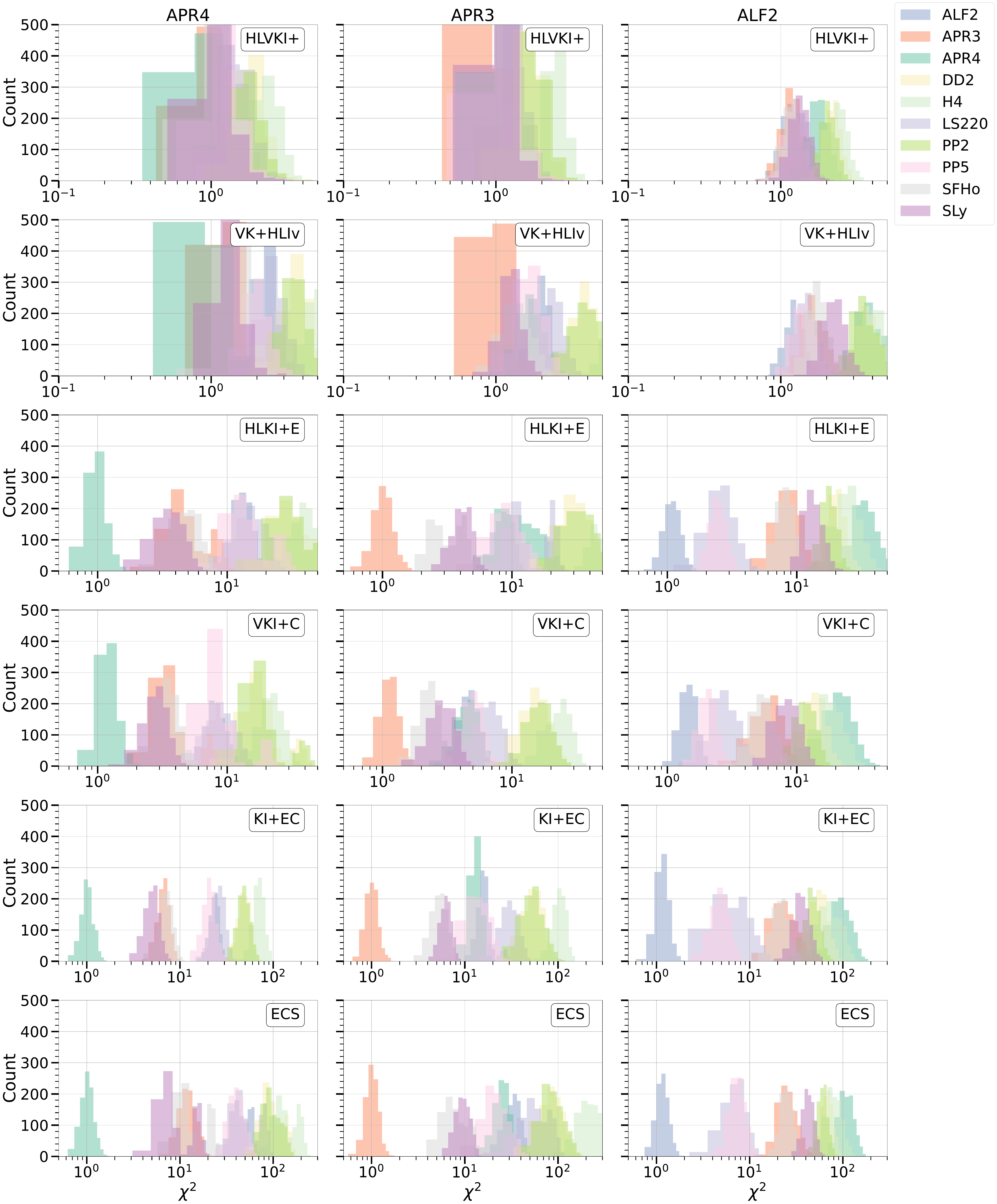}
\caption{Chi-square histograms for 100 events from those 500 with the smallest error in combined tidal deformability. The injected
EOS is listed along the top, and the colored histograms represent the result assuming a second EOS model, including the original injection.
Detector networks are organized by sensitivity row-wise with the most sensitive network at the bottom.  While the peak of the injected EOS histogram is generally recovered with the smallest $\chi^2$ value despite detector sensitivity, in the two networks which do not contain at least one XG detector, the histograms are not distinguishable and we cannot claim that this test is effective in distinguishing EOSs at loud SNRs. However, in networks with at least one XG detector, the correct EOS is consistently recovered with its distribution clearly separated from other EOS models. The same trend is also seen in Fig \ref{fig:chisq-lamt}.}
\label{fig:chisq-SNR}
\end{figure*}

\subsection{Model Selection}
Figures \ref{fig:chisq-lamt}, and \ref{fig:chisq-SNR} show the primary results from our model selection procedure. Here we plot the distribution of the $\chi^2$ statistic defined in Eq.~(\ref{eqn:chisq}) between the observed mass-radius curve and the one predicted by the chosen EOS model. The separation of the distribution for any two EOS signifies the effectiveness of a detector in distinguishing between the injected and test EOS models. In these figures, each row corresponds to a particular detector network, while each column corresponds to a specific injected EOS (label at the top of the column). The $\chi^2$ histograms in each panel are additionally colored to match the EOS color scheme as in Fig.~\ref{fig:EOS_mr}, with the count on the $y$-axis and the $\chi^2$ (in log-scale) on the $x$-axis. 

For detector networks in the top two rows the inferred radius $r^k$ is very different from that predicted radius $r^M$ by any of the models (see top left and middle panels in Figs.~\ref{fig:MR_error_loudest} and \ref{fig:MR_error_bestmeasured}) which would cause $\chi^2$ to be large. However, at the same time, the uncertainties in the measurement ($\sigma_i$) are also large. Consequently, for networks with poorer sensitivity, the $\chi^2$ will tend to be equally small no matter which EOS model the events are compared to.

The story is different when the radius uncertainty $\sigma$ of a detector network is small. For such detectors, the bias-corrected radius differs significantly from the predicted radius found using a model other than the true one, but agrees very well with the predicted radius of the true model. Consequently, the ratio within the sum in Eq.~(\ref{eqn:chisq}) is small only when the set $\{r_i^M\}$ corresponds to the true EOS. This is the reason why the $\chi^2$ distributions for the models other than the true one have far greater values than they are for the true model in the bottom two rows. We find that the method accurately recovers the injected EOS model among a larger set of models than was used for injection. In addition, we have also used a much larger sample of events for our work compared to previous studies \cite{LIGOScientific:2019eut, Ghosh:2021eqv, Pacilio:2021jmq, Biswas:2021pvm}.

We stress that the power of the $\chi^2$ statistic introduced lies in discriminating between the different EOS models when measurement uncertainties are small; with less sensitive detector networks there is no way to distinguish one EOS model from another. The absolute value of the $\chi^2,$ however, has no significance. 

Across different detector networks, when the injected EOS is close in the M-R parameter space to the comparison EOS, the distribution is most often confused with the true EOS as show by the proximity of its histogram to the true one. For example, in the least sensitive detector networks, or top rows of Fig \ref{fig:chisq-SNR} and \ref{fig:chisq-lamt}, the overlap between the resultant three distributions of ALF2, SLy, and PP5 is total, and even with one XG detector, they still overlap significantly. It is only in the best detector networks (bottom two rows) that they begin to become indistinguishable. Meanwhile, comparing ALF2 to APR4 or H4, even in some of the least sensitive networks, their distributions already diverge from the true ALF2 one. This follow from the simple fact that at low mass in the mass-radius curve, ALF2 lies very close to PP5 and at high mass close to SLy and would therefore naturally match more closely with its nearest neighbors while the distance between ALF2 and APR4 or H4 is significant and therefore not well matched (Kashyap et al.~\cite{Kashyap:2022wzr} discuss how distinguishability of EOS models changes with respect to the $L_2$ distance between them).

In general, as the sensitivity of the networks increases, so too does the separation of the posterior distributions. In the lower sensitivity networks from both the highest SNR and best measured tidal deformability data sets, the distributions overlap significantly, and it is only with the inclusion of at least one XG detector that the distributions become at all distinguishable. Across EOS and data sets in networks with at least one XG detector, the smallest $\chi^2$ value always corresponds to the injected EOS and its peak is distinguishable from the EOS with the next smallest $\chi^2$ value. There is not a significant separation of the true EOS from its neighbors, however, until we begin to include at least two XG detectors in the network. In these most sensitive networks, the true EOS centers around one, effectively recovering the EOS, and there are an order of one hundred separations between it and its neighbors, giving hope that XG detector networks may be able to distinguish clearly between these, and other EOSs.

\begin{figure*}
\centering
\includegraphics[width=0.99\textwidth]{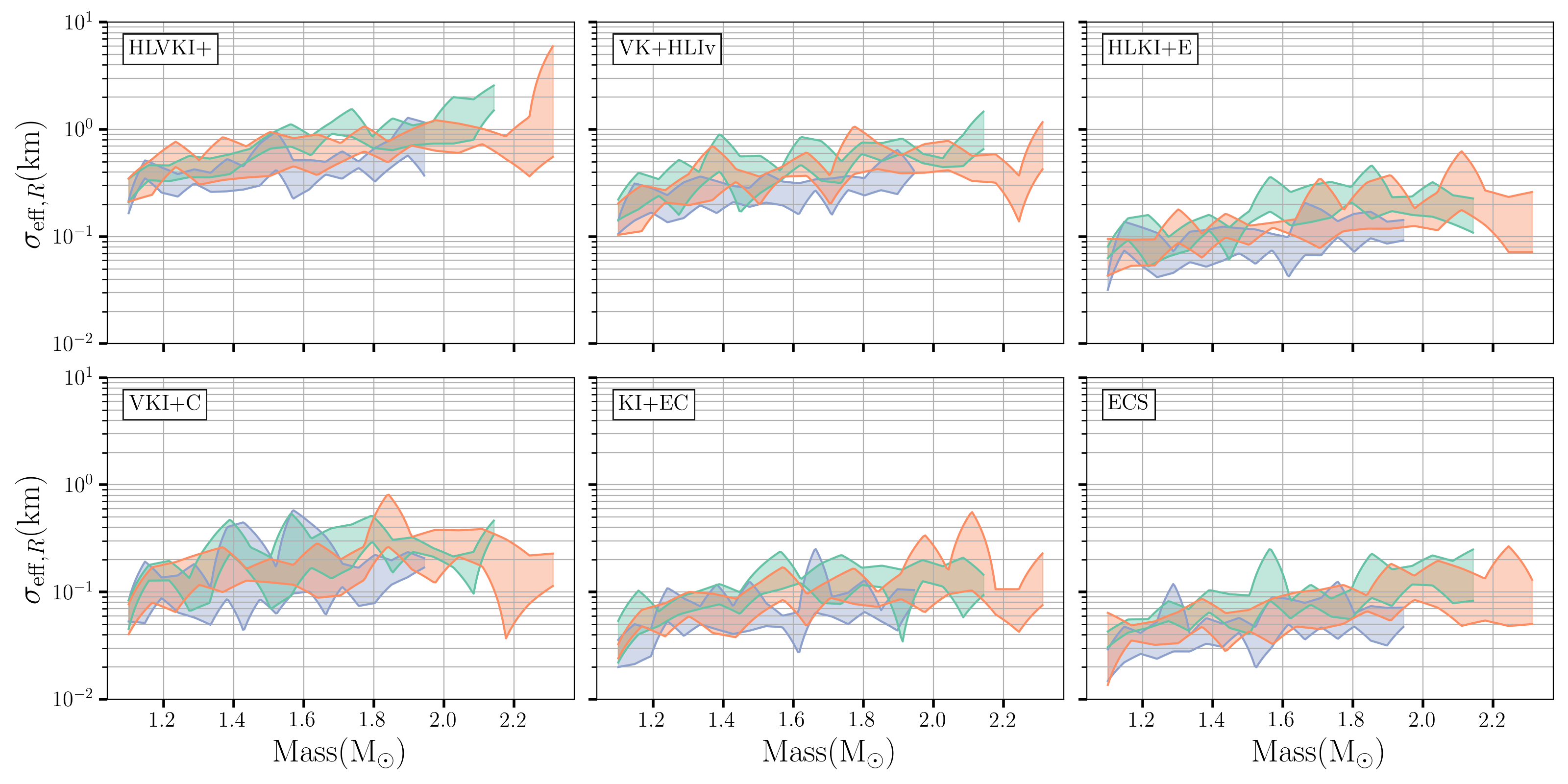} \\
\includegraphics[width=0.99\textwidth]{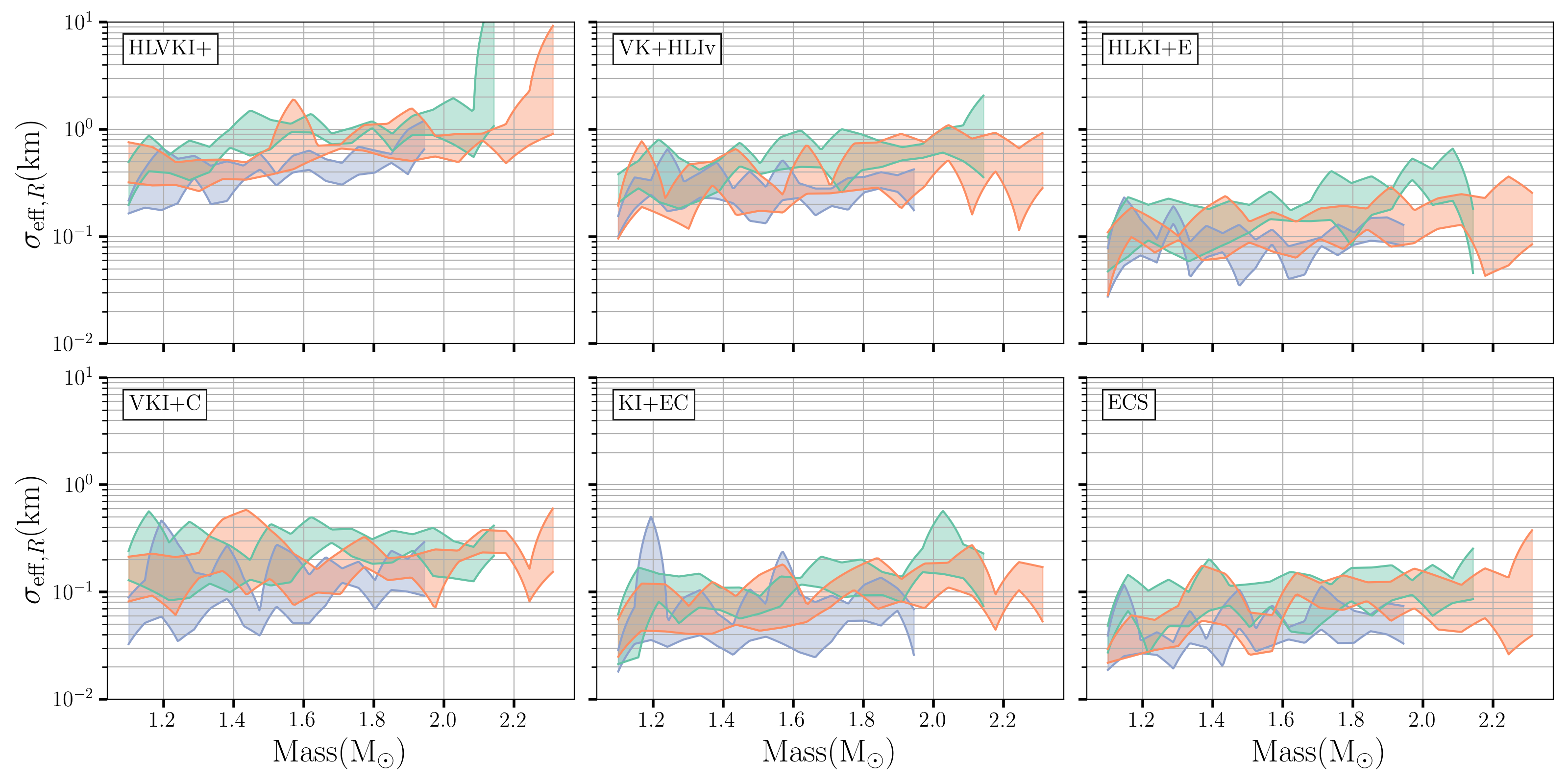}

\caption{Cumulative radius error in each mass bin by square harmonic sum assuming constant radii in each mass bin. The \textit{upper panel} shows the 100 events randomly selected from the 500 events with the best measurement of $\tilde{\Lambda}$ while the \textit{bottom panel} shows the same result for 100 events randomly selected from 500 events with the best SNR. The band for each EOS shows the uncertainty due to random sampling. The color encodes the results for each EOS and is the same as Fig. \ref{fig:EOS_mr}. We find generically that errors in radii are larger for larger masses across detector networks and data sets due to smaller accuracy in the measurement of smaller tidal deformability.}
\label{fig:combraderr}
\end{figure*}

\subsection{Combining Radii Errors from Multiple Events}
In Fig.~\ref{fig:combraderr}, we present the results of combining the radius uncertainties of multiple events binned in individual masses of neutron stars in the range 1.0 M$_\odot$ to the maximum mass supported by the EOS used for the injection, using the method described in Sec.~\ref{sec:comberr}. We plot the  effective errors in the radii of a particular mass bin for three EOS models, with the colors the same as in Fig. \ref{fig:EOS_mr}. The color bands show the variation in the combined error due to bootstrapping while selecting 100 events out of 500 best events according to two different criteria (best SNR and best measured $\tilde{\Lambda}$) as described in the previous sections. We've found that this selection of events does not make a significant difference to the results.

One of the crucial features of these plots is the increase in the  effective radius uncertainty with the increase in the masses of the individual neutron stars. This is again due to the small tidal deformability of heavy neutron stars and poor accuracy in their measurements irrespective of the EOS and the detector network, leading to the poor measurement of radii via the $C$-$\Lambda$ universal relation. Smaller radii and tidal deformabilities at higher masses result in poorer constraints of the EOS at higher densities, which is usually near the neutron star core. As expected, we find an improvement in the radii uncertainties for all mass bins according to the $\sqrt{N}$ law, where $N$ is the number of events combined in each mass bin.

We find the uncertainties to be smaller than 1 km by combining 5 or more events in any mass bins irrespective of the network chosen. The HLVKI+ network has typical errors to be around 1 km for all masses and becomes as large as 3 km even after combining multiple events. The addition of one XG detector to the network improves the radii uncertainties by an order-of-magnitude with a typical value of 100~m, the smallest value of 30~m, and the largest value of 1~km. The best radius measurement, however, is accomplished by combining both ET and CE. We show the results for two such networks of detectors where uncertainties could be as small as 20~m with almost all of the bins having uncertainties smaller than 100~m (i.e., $\sim 1\,\%$). We emphasize again that in these calculations, we use Fisher Information Matrix to approximate the uncertainties, which are a lower bound. We defer the work of accurate analysis using Bayesian Monte Carlo methods to future work.

\subsection{Discussion}
\label{sec:discussion}
\begin{figure*}
\includegraphics[width=1.0\textwidth]{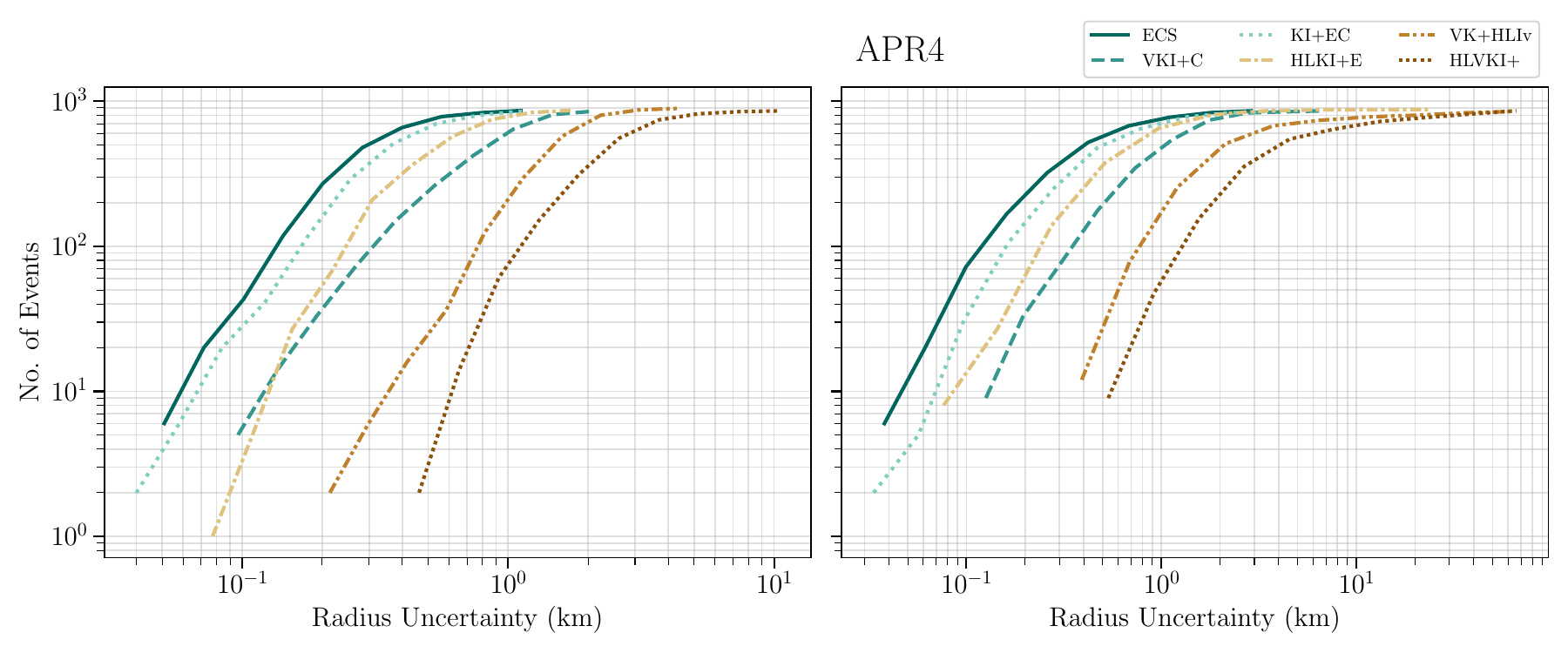}
\includegraphics[width=1.0\textwidth]{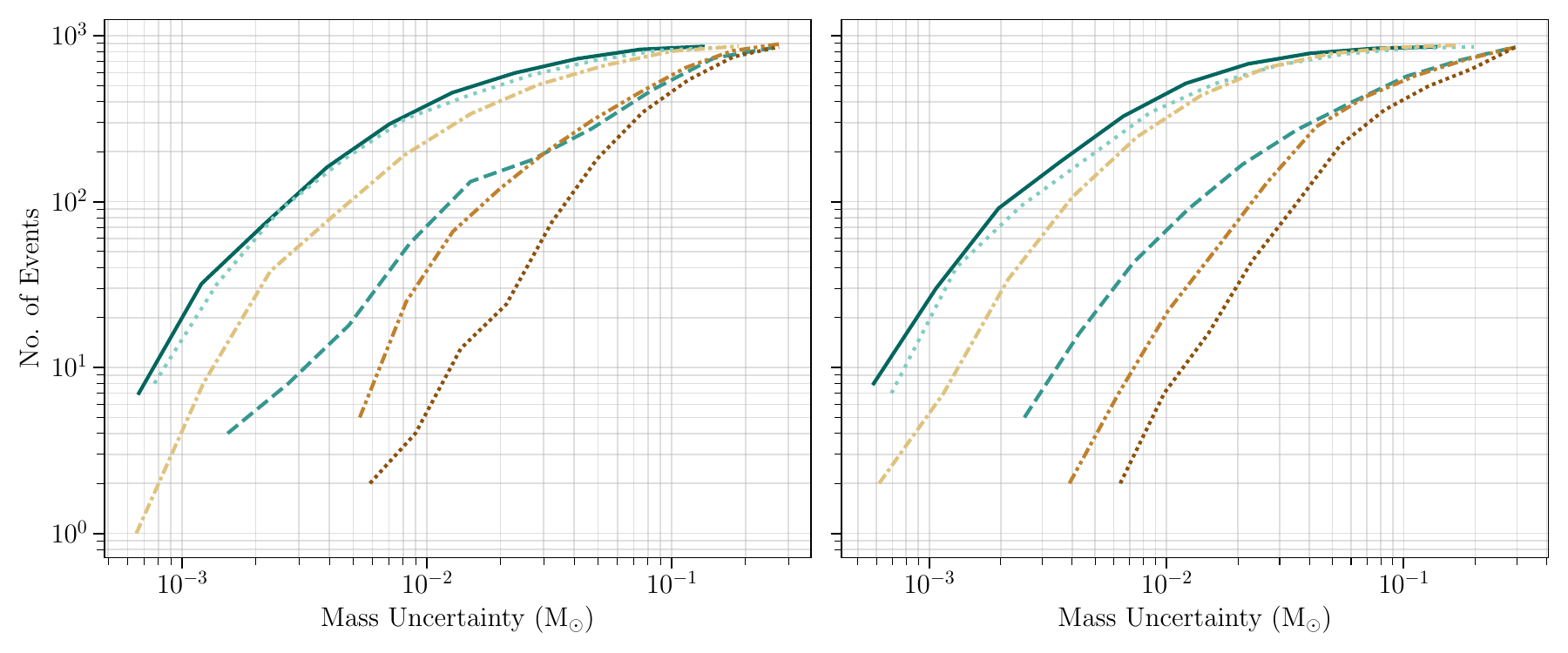}
\caption{\textit{Upper Panel:} Cumulative histograms 
of the uncertainty in neutron star radii in km (top two panels) and masses in solar mass (bottom two panels)  multiplied by the total number (860) of neutron stars in the 430 selected
BNS events. The left panels are for events with the best measured tidal deformability and the right panels are for events with the highest SNRs. The different curves correspond to different detector networks considered in this study. These plots show that even the inclusion of just one XG detector (VKI+C or HLKI+E) leads to a vast improvement in the precision of radii measurements. Such detectors could measure the radius to within about 200~m for several events. A network containing two or three XG detectors would improve by a factor of a few. On the other hand, companion masses are better measured by a network that has ET (0.01 $M_\odot$ to 0.001 $M_\odot$) whose lower frequency performance helps in more accurate determination of the chirp mass and the mass ratio.}
\label{fig:radhistogram}
\end{figure*}

The result of our analysis for the best-SNR and best-measured tidal deformability data sets is promising for networks including XG observatories.  Advanced LIGO and Virgo and their upgrades in the near future are expected to observe tens of events with moderate SNR (i.e., SNR$>40$) and a handful of high-fidelity (SNR$>100$) events over a two-year period (cf.\, \ref{tab:num-events}, columns 2 and 3). Without any XG observatories, the best fractional uncertainty in radii measurements for the top 100 events with best measured tidal deformability is 5--10\%, with more than half above 10\%, as seen in Fig \ref{fig:radhistogram}. This means it will be  difficult for these networks to distinguish between even the most disparate set of EOS models considered in this paper. However, with the inclusion of just one XG detector, the best results show only a 0.8\% uncertainty in radii, with half of the events reporting only 6\% or less, allowing EOS to become partially distinguishable.  Meanwhile, networks with at least two XG detectors tell a completely different story.

In our most sensitive networks, we will be able to measure the radii of neutron star sources to 0.5\%, with half at 3\% or less, as seen in Fig \ref{fig:radhistogram}. However, we have not taken into account the models of the crust of neutron stars which themselves can be 100~m (i.e., ~1\% of the radius), so further work is required to better characterize the meaning of measurement accuracies below this accuracy. These precise measurements, however, result in $\chi^2$ distributions that are easily distinguishable, well separated, and centered for both the loudest SNR and best-measured tidal deformability event sets. Consequently, XG networks will be able to distinguish between different EOS models (even ones that are sufficiently close to each other in $L_2$ measure of distance) and place stringent constraints. Overall, the results of these data sets reveal an avenue for future research that deserves to be pursued further.

\section{Summary and Conclusions}
\label{sec:conclusions}

In this work, we report on the improvements in the inference of the dense matter equation of the state of neutron stars with the current and next-generation gravitational-wave detectors based on their expected design sensitivity curves. We evaluate the measurement uncertainties for hundreds of thousands of events and consequently, it is not possible to carry out a Bayesian inference analysis of the events as that would currently take a formidable amount of time. Instead, we use the Fisher matrix approximation to compute the 1-$\sigma$ uncertainties and correlations of the binary neutron star parameters, including the companion masses and the effective tidal deformability $\tilde\Lambda$ using the \textsc{IMRPhenomPv2NRTidal} waveform. The multivariate distributions of the binary parameters obtained from gravitational-wave observations, together with two universal relations, namely, Eqs.(\ref{eqn:unirel1}) and (\ref{eqn:unirel2}), allow us to infer the mass-radius posteriors of companion neutron stars.   Since the universal relations are not exact, the inferred radii posteriors have systematic biases. We have shown that these systematic biases can be corrected for when comparing the measured mass-radius posteriors with that predicted by a specific equation-of-state model. Our bias-correction method is equivalent to comparing the model mass-tidal deformability predictions directly with the gravitational-wave data but computationally inexpensive since bias corrections are known a priori and don't need to be generated on the fly. Moreover, the method avoids having to repeat the likelihood calculations and computations of posteriors for every plausible equation-of-state.

We employed this new method to compare three disparate model equations of state with simulated gravitational-wave measurements for assuming the true equation of state to be one of the 10 models. Our results demonstrate that the method can uniquely identify the correct equation-of-state when the detector network contains at least one XG observatory (either Einstein Telescope or Cosmic Explorer).  It will be difficult to distinguish between different plausible equations of state with the current network of LIGO, Virgo and KAGRA observatories or their proposed improvements (A+ or Voyager). However, with the addition of at least one XG observatory, it will be possible to draw firm conclusions about the true equation-of-state describing dense matter in neutron star cores. Moreover, we find vast improvements in the measurement uncertainties of neutron star radii with two or more next-generation observatories in the network. More specifically, we find that radius uncertainties are a few hundred meters for networks with one or more next-generation observatories, while this would be 1 km in a network with the LIGO-Virgo-KAGRA network and their future upgrades. However, we found that the overall accuracy of radii measurements decreases with increasing neutron star mass. This is because tidal deformabilities are smaller and more difficult to measure for more massive neutron stars.

Building more sensitive gravitational-wave observatories is crucial to constraining plausible EOS models---measurements that can inform not only the gravitational-wave community but also the nuclear physics and astronomy communities at large. In this light, the radius of a typical NS can be constrained to better than 30 m, at the lower end of the expected range of neutron star masses, with joint detections of events over two years in Einstein Telescope and Cosmic Explorer. 

\section*{Acknowledgements} \label{sec:acknowledgements} We thank the Cosmic Explorer Team members and the \textsc{Matter Group} of the LIGO Scientific Collaboration for discussions and feedback. RH, AD, and BSS were supported by NSF grant numbers PHY-2012083, AST-2006384 and PHY-2207638.  SB acknowledges support from the Deutsche Forschungsgemeinschaft, DFG, Project MEMI number BE 6301/2-1.

\appendix
\section{Miscellaneous plots}
\label{sec:appendix}
\begin{figure*}
    \centering
    \includegraphics[width=0.99\textwidth]{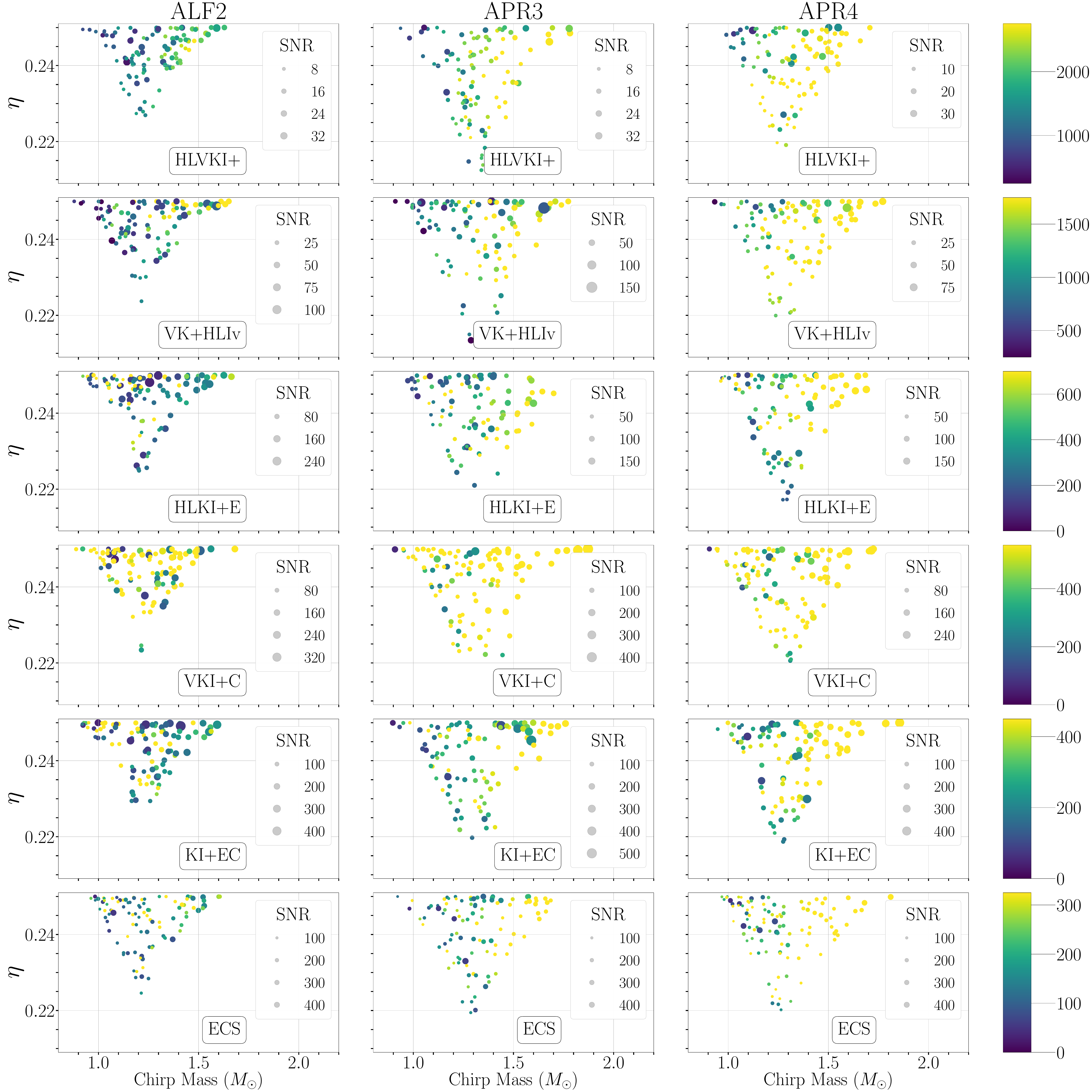}
    \caption{The plot shows the radius error (in color) for 100 events with the smallest error in the combined tidal deformability as a function of the symmetric mass ratio and chirp mass. Results are shown for the six detector networks (labelled in each panel) and and three different EOSs (labelled at the top of each column). We recover an approximate trend of increasing radii error with increasing chirp mass and symmetric mass ratio.}
    \label{fig:radius_chirpmass_eta_snr_bestLambdaTilde}
\end{figure*}

\begin{figure*}
    \includegraphics[width=0.99\textwidth]{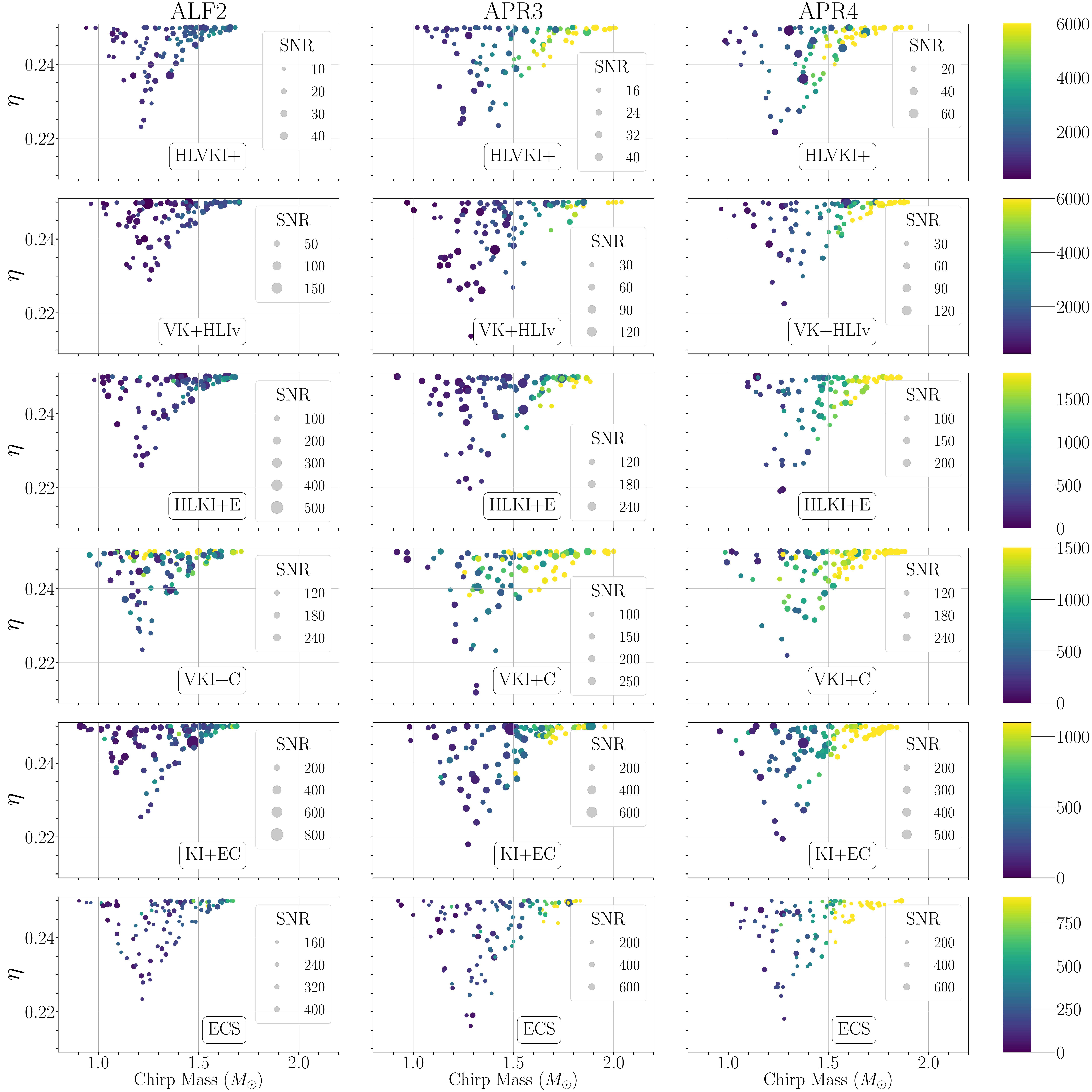}
    \caption{Radius error (in color) for 100 random events out of the 500 which are loudest in SNR for our six 3G detectors and three EOS of choice: left ALF2, middle APR3, right APR4. Detectors are ordered top to bottom as follows: ESa4cCa4c, KI+ECa4c, VKI+Ca4c, HLKI+E, VK+HLIvc, HLVKI+. We see the same trend of increasing radii error with increasing chirp mass and symmetric mass ratio as seen in Fig \ref{fig:radius_chirpmass_eta_snr_bestLambdaTilde}, but it appears more clearly in this data set, especially in increasing chirp mass.}
    \label{fig:radius_chirpmass_eta_snr_loudest}
\end{figure*}

\begin{figure*}
    \includegraphics[width=0.99\textwidth]{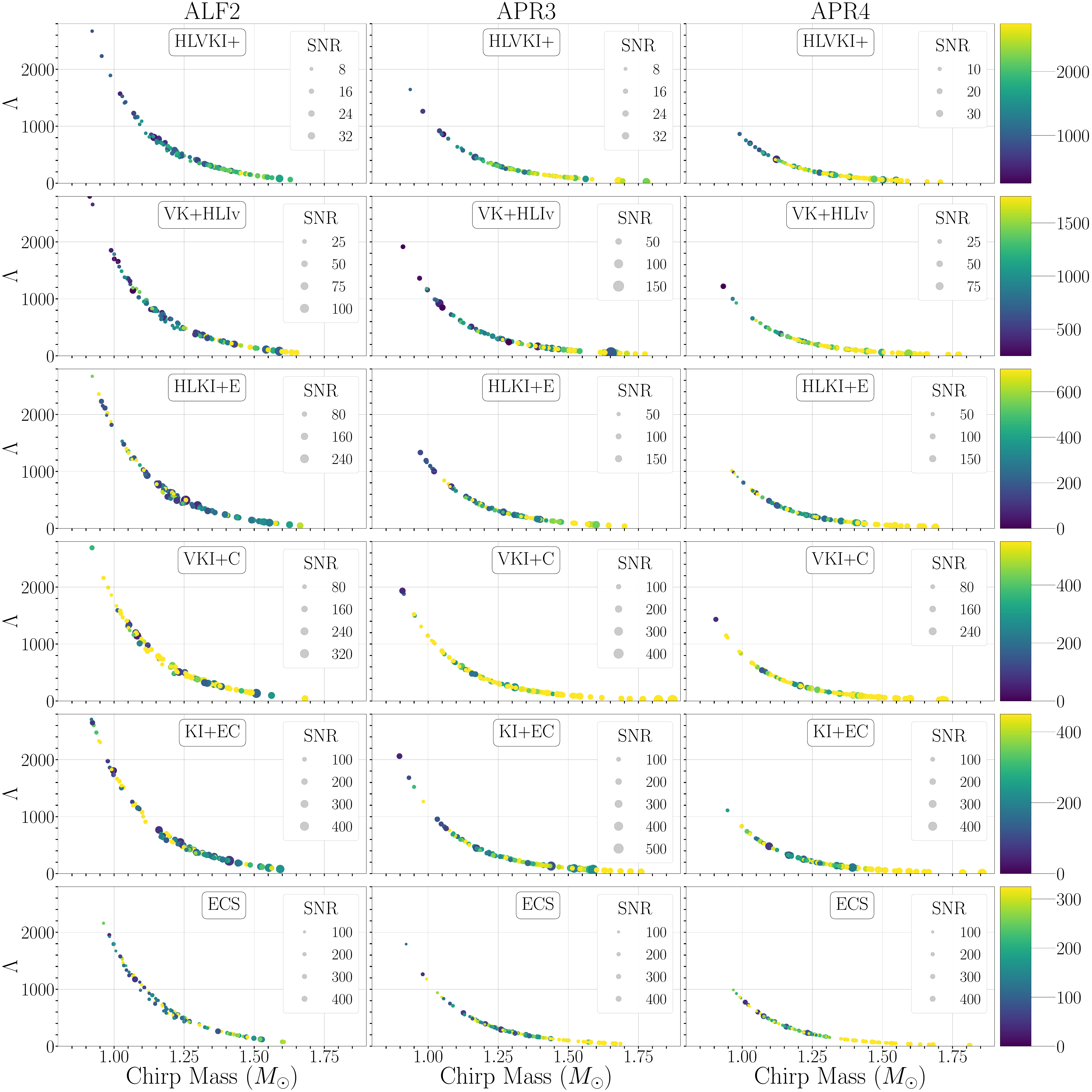}
    \caption{Radius errors (in m) are shown in color for 100 random events out of the 500 which are best measured in tidal deformability for our six 3G detectors and three EOS of choice: left ALF2, middle APR3, right APR4. Detectors are ordered top to bottom as follows: ESa4cCa4c, KI+ECa4c, VKI+Ca4c, HLKI+E, VK+HLIvc, HLVKI+. The relative error in radius naturally decreases for systems with larger combined tidal deformability, and again we see larger radii errors at higher chirp mass. Additionally, we naturally see the largest radii errors and smallest SNRs in near-future detector networks, and significantly better ones in networks with XG detectors.}
    \label{fig:radius_Lambda_chirpmass_snr_bestLambdaTilde}
\end{figure*}

\begin{figure*}
    \includegraphics[width=0.99\textwidth]{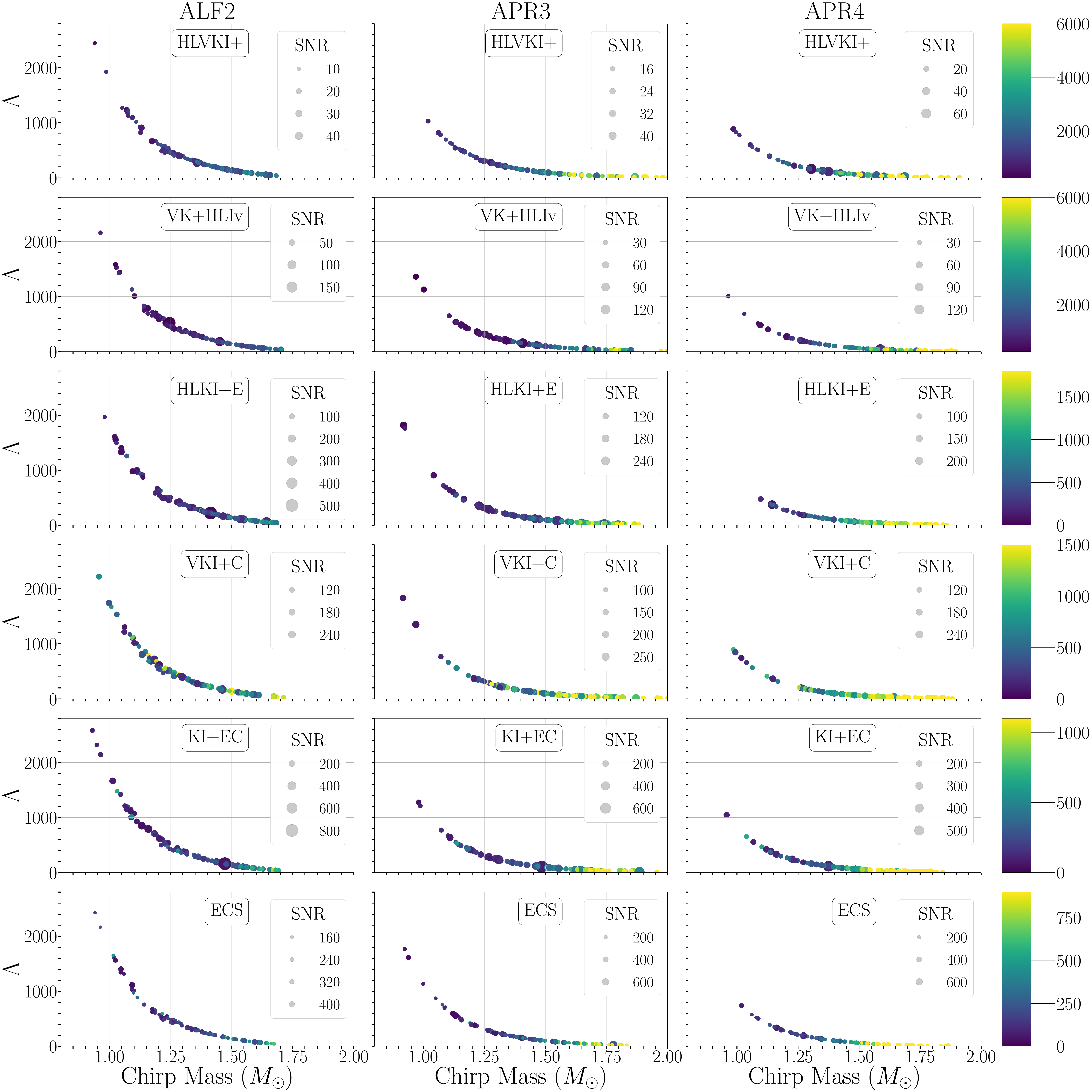}
    \caption{Radius error (in color) for 100 random events out of the 500 which are loudest in SNR for our six 3G detectors and three EOS of choice: left ALF2, middle APR3, right APR4. Detectors are ordered top to bottom as follows: ESa4cCa4c, KI+ECa4c, VKI+Ca4c, HLKI+E, VK+HLIvc, HLVKI+. The same trends appear here as in Fig \ref{fig:radius_Lambda_chirpmass_snr_bestLambdaTilde}. Radius error increases with chrip mass, and decreases with combined tidal deformability. Additionally, as the detector networks themselves improve, we see clear improvements radius error and SNR.}
    \label{fig:radius_Lambda_chirpmass_snr_loudest}
\end{figure*}

In this section we assemble a list of four additional plots to gain a better understanding of the results presented in the main body of this paper. 
These plots show the measurement uncertainty in radius either as a function of chirp mass and symmetric mass ratio in 
Figs.\,\ref{fig:radius_chirpmass_eta_snr_loudest} (for 100 randomly chosen events out of the 500 loudest events) and \ref{fig:radius_chirpmass_eta_snr_bestLambdaTilde} (for 100 randomly chosen events out of 500 events with the best measured tidal deformability) 
or as a function of the chirp mass and tidal deformability in
Figs.\,\ref{fig:radius_Lambda_chirpmass_snr_loudest} (for 100 randomly chosen events out of the 500 loudest events) and \ref{fig:radius_Lambda_chirpmass_snr_bestLambdaTilde} (for 100 randomly chosen events out of 500 events with the best measured tidal deformability). 



\bibliographystyle{apsrev4-1}
\bibliography{refs}

\end{document}